\documentclass[runningheads]{llncs}
\usepackage[T1]{fontenc}
\usepackage{subcaption}
\usepackage{hyperref}
\usepackage{algorithm2e}
\usepackage{algpseudocode}
\usepackage{standalone}
\usepackage{amsmath}
\usepackage{amssymb}
\usepackage{siunitx}

\usepackage{color}
\usepackage[dvipsnames]{xcolor}

\usepackage{graphicx}
\usepackage{color}

\begin{document}
\title{A Holistic Architecture for Monitoring and Optimization of Robust Multi-Agent Path Finding Plan Execution}
\titlerunning{Monitoring and Optimization of Robust MAPF Plan Execution}
%
\author{David Zahrádka\inst{1,2}\orcidID{0000-0002-7380-8495} \and
Denisa Mužíková\inst{1}\orcidID{0009-0000-8535-8439} \and
David Woller\inst{2}\orcidID{0000-0001-8809-3587} \and
Miroslav Kulich\inst{2}\orcidID{0000-0002-0997-5889} \and
Jiří Švancara\inst{3}\orcidID{0000-0002-6275-6773} \and
Roman Barták\inst{3}\orcidID{0000-0002-6717-8175}}

\authorrunning{D. Zahrádka et al.}
%
\institute{Faculty of Electrical Engineering, Czech Technical University in Prague, Prague, Czech Republic \email{muzikden@fel.cvut.cz}\\
\and
Czech Institute of Informatics, Robotics and Cybernetics, Czech Technical University in Prague, Prague, Czech Republic \email{\{david.zahradka,david.woller,kulich\}@cvut.cz}\\
\and
Faculty of Mathematics and Physics, Charles University, Prague, Czech Republic
\email{\{svancara,bartak\}@ktiml.mff.cuni.cz}}

\maketitle              
%
\begin{abstract}
The goal of Multi-Agent Path Finding~(MAPF) is to find a set of paths for a fleet of agents moving in a shared environment such that the agents reach their goals without colliding with each other.
In practice, some of the robots executing the plan may get delayed, which can introduce collision risk.
Although robust execution methods are used to ensure safety even in the presence of delays, the delays may still have a significant impact on the duration of the execution.
At some point, the accumulated delays may become significant enough that instead of continuing with the execution of the original plan, even if it was optimal, there may now exist an alternate plan which will lead to a shorter execution.
However, the problem is how to decide when to search for the alternate plan, since it is a costly procedure.
In this paper, we propose a holistic architecture for robust execution of MAPF plans, its monitoring and optimization.
We exploit a robust execution method called Action Dependency Graph to maintain an estimate of the expected execution duration during the plan's execution.
This estimate is used to predict the potential that finding an alternate plan would lead to shorter execution.
We empirically evaluate the architecture in experiments in a real-time simulator which we designed to mimic our real-life demonstrator of an autonomous warehouse robotic fleet. 


\end{abstract}
\section{Introduction}
\setcounter{footnote}{0}

Coordinating a fleet of mobile agents is an important problem with practical applications in autonomous warehouses~\cite{lehoux-lebacque2024multiagent,Wurman}, airfield operations~\cite{Morris}, and even traffic junction coordination~\cite{Dresner}.
\emph{Multi-Agent Path Finding}~(MAPF) is an abstract model of these coordination problems, where the goal is to find collision-free paths for agents moving in a shared environment represented as a graph.
The problem of finding an optimal MAPF solution has been shown to be NP-hard for a wide range of objective functions~\cite{surynek2010optimization}.
MAPF methods can be divided into two classes, which differ in what guarantees they provide on the cost of the solution.
The methods in the first class do not guarantee optimality (or even completeness), but are computationally non-demanding and scale well to a large number of agents, for example, WHCA*~\cite{silver2005cooperative}, SIPP~\cite{phillips2011sipp}, MAPF-LNS~\cite{li2021anytime} and LaCAM*~\cite{Okumura2023}.
Provably optimal and bounded suboptimal algorithms form the second class. Prominent optimal algorithms are, for example, Conflict-Based Search~(CBS)~\cite{sharon2015conflict,li2019symmetrybreaking}, or approaches reducing MAPF into Boolean Satisfiability problem~(SAT)~\cite{Bartak}, Answer Set Programming~(ASP)~\cite{erdem2013general}, and Mixed Integer Programming~(MIP)~\cite{JYU}.
Bounded suboptimal solvers are often based on optimal solvers, such as Enhanced CBS~(ECBS)~\cite{barer2014suboptimal} and its new variant Explicit Estimation CBS~(EECBS)~\cite{li2021eecbs}, or a bounded suboptimal SAT-based algorithm~\cite{surynek2020bounded}.

Executing MAPF plans with real robots poses more challenges.
Individual robots may get delayed due to numerous unexpected circumstances, such as mechanical problems, uncertainties in localization or control, or even due to the presence of unmodeled agents in the shared environment, e.g., humans.
These delays lead to a desynchronization of the robotic fleet, and as a consequence, the robots may collide with each other or get stuck.

Collisions and deadlocks can be prevented by finding the so-called \emph{k-robust plans}, which are collision-free even if any number of the agents are delayed by up to \emph{k} time steps~\cite{Atzmon2020,Chen2021}.
However, this comes with the disadvantage of increased solution costs.
A more realistic approach is presented in~\cite{Ma2017}, where it is assumed that each individual action may be delayed with some probability.
This information is used within a CBS-based solver to minimize the approximate \emph{expected makespan}, and it is shown that this leads to a reduced real makespan of executed plans compared to standard solvers. 
Wagner et al.~\cite{Wagner2022} introduce the Space-Level CBS, an ECBS-based planner that provably reduces the number of crossings between paths of different agents to minimize the number of possible interactions between them.
In~\cite{Wu2024}, a Space-Order CBS that builds upon Space-Level CBS~\cite{Wagner2022} is presented.
It directly plans a temporal graph, and thus minimizes the number of coordination actions explicitly.

However, as long as there is a risk of interactions between agents, none of these plans is guaranteed to be perfectly executed with real-life robots.
Ensuring a deadlock-free and collision-free execution requires the use of \emph{robust execution methods}, which use \emph{retiming} to change the time at which actions are executed to ensure safe execution.
The first such method is MAPF-POST~\cite{Hoenig2016}, which post-processes a plan produced by a MAPF solver and builds a graph called a \emph{temporal network}.
Its nodes represent events corresponding to an agent entering a location, and its edges represent temporal precedences between different events.
The graph is augmented by additional vertices that are used to provide a guaranteed safety margin between the agents, and the authors proved that MAPF-POST ensures collision-free and deadlock-free execution.
Gregoire et al.~\cite{Gregoire2017} introduce a robust multi-robot trajectory tracking method RMTRACK.
The ideas are similar to~\cite{Hoenig2016}, but with a key difference that RMTRACK works in a coordination space and it ensures that the realized trajectory remains in the same homotopy class as the planned trajectory.
The Action Dependency Graph~(ADG)~\cite{Honig2019} is a temporal network -- similarly to MAPF-POST -- but temporal precedence constraints are expressed between actions rather than events at locations.
This makes the graph simpler with less communication among agents and has stronger guarantees on collision-free operation.

While the techniques mentioned above guarantee safe execution, the final quality of the realized trajectories can substantially degrade in the presence of delays.
This can be solved by either \emph{rescheduling} of the agent's actions, which means switching the order in which the agents pass through a given location, or \emph{replanning}, which finds a completely new set of paths for the agents.
Replanning methods are more powerful; they are also capable of only rescheduling if they find the same paths but with different ordering of the agent on crossroads and any standard MAPF solver can be used.
However, replanning is also more costly, because of the larger search space.

Many recent approaches focus on methods for rescheduling.
In~\cite{Berndt2020,Berndt2024}, a Switchable ADG~(SADG) is presented, which creates an inverted version of each dependence between actions of different agents.
Rescheduling in the form of deciding whether to use the original or the reverse dependence for each action is then formulated as a mixed integer linear program that is solved in a closed-loop feedback scheme.
Similarly, in~\cite{zahradka2023solving}, rescheduling is formulated as a modification of the Job Shop Scheduling problem, which is then solved using a Variable Neighborhood Search metaheuristic.
Another approach to repair the plan is to introduce additional delays, as presented in~\cite{Kottinger2024}.
The plan repair problem is formulated as an optimization problem with the goal of finding a minimal number of additional delays required to repair the plan. 
To solve the problem, any arbitrary MAPF solver can be used on a reduced graph built as a superposition of the original paths. 
This approach is substantially faster than replanning from scratch.
Finally, in~\cite{Feng2024}, a Switchable Edge Search is proposed, which aims to find a new temporal graph based on the given SADG making use of an A*-style algorithm.

The disadvantage of the above-mentioned approaches is that they run the optimization \emph{during} execution.
To avoid this, a Bidirectional Temporal Plan Graph~(BTPG)~\cite{Su2024} is proposed, which allows switching between different orderings of agents during execution based on a "first-come-first-served" manner.
These different orderings are built \emph{before} the execution.
This is done by consecutively checking whether edges indicating dependencies between actions of different agents can be transformed into a bidirectional pair without causing an oriented cycle in the BTPG. 

Despite the enormous recent activity in finding a robust planning \& execution framework for MAPF, several open questions still need to be answered.
One of the most interesting open questions is when to reschedule or replan and whether rescheduling still provides a good quality solution or whether replanning would be worth the added computational cost.

In this paper, we present a holistic architecture for robust execution of MAPF plans, monitoring of the execution's progress and its optimization.
With this architecture, both robust execution and monitoring are performed in the robust execution layer, and it also supports the integration of methods to determine whether an alternate plan could be beneficial.
We present one such method that is capable of estimating the impact of currently accumulated delays by computing the so-called \emph{slack} and using it to decide whether an alternate plan has the potential to decrease the duration of the execution.
We find alternate plans to demonstrate the effectiveness of the method by replanning.
This holistic architecture and the method were originally proposed in~\cite{zahradka2025holistic}.
In this paper, we extend our previous work in three ways. 
First, we present a real-life robotic fleet demonstrator, which we use to experimentally verify our methods.
Then, we present a robotic fleet simulator that we developed for the purposes of evaluating algorithms for execution optimization on a large scale.
The architecture of the simulator closely follows the design of our experimentally verified real-life demonstrator.
The simulator also uses a realistic model of an unmodeled agent to measure the effect caused by unexpected delays.
Finally, we present an experimental evaluation of the holistic framework described in~\cite{zahradka2025holistic}.
We use an improved version of the original slack-based predictive replanning method.
The experiments were carried out on multiple benchmark maps and with larger fleets of varying sizes in $1300$ experiments in total. 

\section{Background}
In this section, we first formulate MAPF and the concept of $1$-robust solutions, which are used throughout this paper.
We show examples of a MAPF instance, a plan, and define relevant optimization criteria.
Then, we describe ADG, which is a MAPF robust execution method on which we demonstrate the holistic architecture for robust execution, its monitoring and optimization.

\subsection{Multi-Agent Path Finding}
We use the same notation as in~\cite{zahradka2025holistic}.
A MAPF instance \emph{M} is a pair $M = (G, A)$, where \emph{G} is a graph $G = (V,E)$ representing the shared environment and $A$ is a set of mobile agents.
Each agent $k\in A$ has a starting location $s_k \in V$ and a goal location $g_k \in V$.

\begin{figure}[htb]
    \centering
    \begin{subfigure}[c]{0.45\linewidth}
       \includegraphics[width=1.0\textwidth]{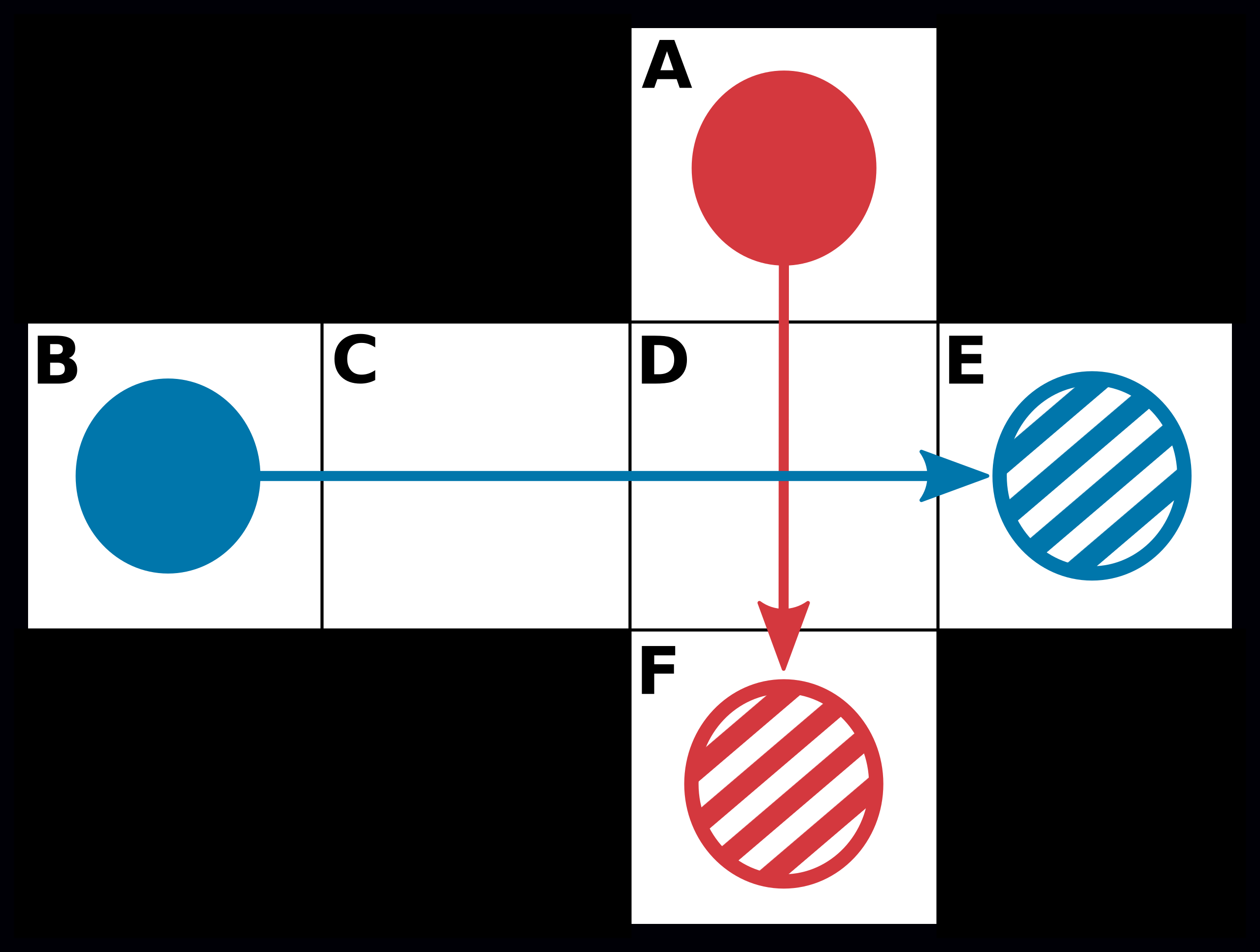} 
       \caption{}
       \label{fig:small_mapf_example_solution}
    \end{subfigure}
    \begin{subfigure}[c]{0.45\linewidth}
       \includegraphics[width=1.0\textwidth]{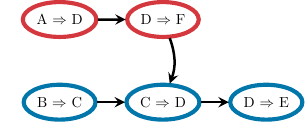}
       \caption{}
       \label{fig:small_mapf_example_adg}
    \end{subfigure}
    \caption{Example MAPF plan (a) and its ADG (b). The filled circles are agents and the hatched circles are their goals. Example plan sourced from~\cite{zahradka2025holistic}.}
    \label{fig:small_mapf_example}
\end{figure}

The task is to find a plan $\pi_k$ for each agent $k \in A$ that corresponds to a path starting in $s_k$ and ending in $g_k$.
Time is discretized into timesteps. 
In each timestep $i$, an agent can either wait in its current vertex or move to a neighboring vertex over an edge $e \in E$.
Therefore, each plan $\pi_k$ can be considered a sequence of \emph{move} and \emph{wait} actions.
After executing its plan, the agent $k$ remains occupying its goal location $g_k$.
An example plan can be seen in Fig.~\ref{fig:small_mapf_example_solution}.

Furthermore, we require that each pair of plans $\pi_k, \pi_{k'}; k \neq k'$ is collision-free (commonly called conflict-free). 
Multiple types of conflicts exist~\cite{stern2019multi}.
We forbid the so-called \emph{vertex conflicts}, where two agents may not occupy the same vertex at the same timestep $i$.
We also prohibit the so-called \emph{swap conflicts}, where two agents travel across the same edge at the same time in opposite directions.
The remaining types of conflict that we prohibit are the \emph{following} and \emph{cycle conflicts}.
The following conflict occurs when an agent enters a vertex at the same timestep that another agent left it.
A cycle conflict describes a situation in which there is a group of agents moving in a circular pattern, each agent following one unique agent in the group.

Solutions that do not contain any of the aforementioned conflicts are called $1$-robust solutions~\cite{Atzmon2020}.
This means that an agent may not enter any $v \in V$ at timestep $i$ if $v$ was occupied in the previous timestep $i-1$ by another agent $a'; a \neq a'$.
Adhering to this rule prevents all \emph{swap}, \emph{following}, and \emph{cycle} conflicts.

The quality of the plan is measured by \emph{optimization criteria}, such as \emph{makespan} and \emph{flowtime} (also known as \emph{Sum of Costs}, or SOC).
SOC is given as a sum of the lengths of all agents' plans, while makespan is defined as the length of the longest plan.
The same criteria can be used to measure the quality of the plan's \emph{execution} by real or simulated agents.
SOC can be measured as the sum of completion times of agents' last actions, while makespan can be measured as the time when the last agent completed its last action.
In this work, we focus on measuring \emph{SOC}, but our methods are not exclusive to this objective function.

\subsection{Action Dependency Graph}
The Action Dependency Graph~(ADG)~\cite{Honig2019} is a directed acyclic graph $G_{ADG} = (V_{ADG}, E_{ADG})$, where $V_{ADG}$ are the actions in the agents' plans and $E_{ADG}$ are directed edges representing temporal dependencies of different actions. 
Each action $a_i^k \in V_{ADG}$ is a pair $(p_i, p_{i'})$ and represents that an agent $k$ in timestep $i$ starts moving from position (vertex) $p_i$ and arrives in position $p_{i'}$ at timestep $i' = i+1$.

If there is an edge between a pair of actions $a_i^k, a_{i'}^{k'}$ -- in other words, $(a_i^k, a_{i'}^{k'}) \in E_{ADG}$ -- there is a temporal precedence between them: agent $k'$ can start executing $a_{i'}^{k'}$ only after the agent $k$ has finished executing $a_{i}^{k}$.

There are two types of such edges: Type 1 edges ($E^1 \subset E_{ADG}$) represent temporal precedences within the path of a single agent: $(a_{i}^k, a_{i'}^{k})$.
They represent that an agent $k$ may start moving from some position only after the agent finished moving into it. 
Type 2 edges ($E^2 \subset E_{ADG}$) indicate dependencies between actions of different agents: $(a_{i}^{k}, a_{i'}^{k'})$.
A Type 2 edge represents that an agent $k'$ can start moving into some shared vertex only after the agent $k$, scheduled to be there earlier, finished moving out of it.
An example ADG can be seen in Fig.~\ref{fig:small_mapf_example_adg}.

The ADG is used for robust execution of MAPF plans.
During execution of a plan, the agents report to ADG whenever they finish an action.
ADG keeps track of completed actions, and whenever an action is finished, ADG marks all outgoing edges of the corresponding vertex as completed.
Once an action has all its incoming edges marked as completed, it is assigned to an agent.
Therefore, the agents may not start executing any action that does not already have all prerequisites fulfilled.
This process is called \emph{retiming}, because it changes the time at which actions are executed depending on the circumstances. 

ADG can ensure safe execution of any plan that does not contain \emph{vertex}, \emph{edge} or \emph{cycle} conflicts even if any agent is arbitrarily delayed, as long as the delay is finite.
However, \emph{following} conflicts lead to an increased execution times due to the nature of $E^2$.
Consider two actions $a_{i}^{k} = (p_m, p_n); a_{i}^{k'} = (p_o, p_m)$.
That is, in timestep $i$, agent $k$ is moving out of $p_m$, and in the same timestep, agent $k'$ is moving into $p_m$.
A Type 2 edge leads from $a_{i}^{k}$ to $a_{i}^{k'}$.
Although agents were assumed to be moving at the same timestep $i$, $a_{i}^{k'}$ may start being executed only after $a_{i}^{k}$ is finished.
For this reason, we also require that no plan contains a following conflicts, or, in other words, that all solutions must be $1$-robust.

\section{Execution Architecture}
In~\cite{zahradka2025holistic}, we proposed a \emph{holistic architecture} capable of i) robust execution; ii) monitoring of execution progress, and iii) deciding whether the fleet encountered delays of sufficient severity that an intervention in the form of rescheduling or replanning could be beneficial.
Another benefit of this architecture is that every step is performed during execution on the robust execution layer.
In this work, we follow the principle and implement it into practice.

\subsection{Execution Monitoring}
Although agents accumulate delays during plan execution, the safety of the plan that is being executed is not endangered as long as a robust execution method is used.
However, delays may progressively desynchronize the agents, increasing the execution's cost.
There are two ways in which the delay of a single agent can contribute to an increased execution cost.
The first is the impact the delay has on the duration of the execution of the agent's own plan.
This increases the execution sum of costs, and if the delayed agent is the one with the longest plan, it also increases execution makespan.

The second type of influence is due to the interactions between the agents.
If there is a vertex through which two or more agents pass -- a \emph{crossroads} -- the MAPF plan specifies the exact order in which the agents move through.
This ordering is represented by the Type 2 edges in $G_{ADG}$.
Due to the temporal constraint, if an agent that is scheduled to pass through earlier is delayed, it may in turn delay other agents, which have to wait until the first agent moves through the crossroads.
In extreme cases, this may result in a cascade that includes the entire robotic fleet, increasing the impact of even a single delayed action.

\begin{algorithm}
\DontPrintSemicolon
\LinesNumbered
\SetKwInOut{Input}{input}\SetKwInOut{Output}{output}
\Input{$G_{ADG}$}
\Output{$G_{ADG}$ with computed $\hat{t}_s$ and $\hat{t}_c$}

\Begin{
\For{$i = 0 $ to $T_{max}$}{
    $V(i) \gets \text{all actions at time step $i$}$\;
    \For{$a' \in V(i)$}{
        $A^P = \{a : \forall a \in V_{ADG} : e(a, a') \in E_{ADG} \}$\;
        \uIf{$A^P == \emptyset$}{
            $\hat{t}_s(a') = i$\;
        }
        \Else{
            $\hat{t}_s(a') = max_{a \in A^P}(\hat{t}_c(a))$\;
        }
        $\hat{t}_c(a') = \hat{t}_s(a') + \hat{t}_x(a')$\;
    }
}
}
\caption{Computing estimates of action start times $\hat{t}_s$ and completion times $\hat{t}_c$.}
\label{alg:time_predictions}
\end{algorithm}

\begin{figure}
    \centering
    \includegraphics[width=0.55\linewidth]{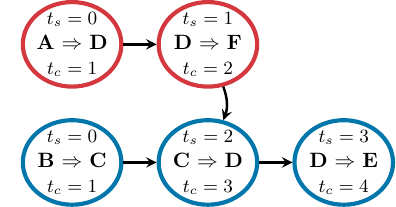}
    \caption{ADG with estimated starting times $t_s(a)$ and completion times $t_c(a)$.}
    \label{fig:adg_prediction}
\end{figure}

The duration of the plan's execution -- its \emph{cost} -- can be estimated by exploiting the $G_{ADG}$. 
The procedure is described in Algorithm~\ref{alg:time_predictions}.
We start with the first timestep $i=0$ of the plan, which contains the set of actions $V(0)$ that the agents will start executing at timestep $0$.
In this paper, we assume that all agents have their initial actions starting at $i=0$, and therefore, $V(0)$ contains exactly one action for each agent, but the procedure can also work with heterogeneous starting times.
For every first action $a_0 \in V(0)$ of every agent $k$, which is scheduled to start at timestep $0$, we set its expected starting time $\hat{t}_s(a_0^k) = 0$.
Then we model the agent's movement capabilities to obtain $\hat{t}_x(a_0^k)$, which is an estimate of the time it takes the agent $k$ to execute $a_0$.
The expected completion time $\hat{t}_c(a_0^k)$ of action $a_0^k$ is then equal to the sum of its estimated start time and its estimated execution time: $\hat{t}_c(a_0^k) = \hat{t}_s(a_0^k) + \hat{t}_x(a_0^k)$.
The corresponding vertex in $V_{ADG}$ is augmented with this value, and we proceed with the next agent.
After we augment every vertex representing an initial action of some agent, we move to the next timestep.

Since we assume starting time $i=0$, every further action $a'$ in timesteps $i \geq 1$ can start only after all preceding actions are finished.
This dependency is represented by the edges $E_{ADG}$ of $G_{ADG}$.
Therefore, we compute the start time of each $a'$ as: $\hat{t}_s(a') = max(\hat{t}_c(a): \exists (a, a') \in E_{ADG})$.
Afterwards, we again use the agent's movement model to compute the estimated completion time of $a'$: $\hat{t}_c(a') = \hat{t}_s(a') + \hat{t}_x(a')$.

This procedure repeats for all subsequent actions in future timesteps. 
If there were some agent $k'$ with the first action in its plan starting in a later timestep $i \geq 1$, we would simply assign $\hat{t}_s(a_0^{k'}) = i$ and proceed as before, although such cases are not considered in this paper.
In the end, we have the estimated completion time for every action in the plan $\hat{t}_c(a) : \forall a \in V_{ADG}$.
An ADG with estimated completion times of actions can be seen in Fig.~\ref{fig:adg_prediction}.

Afterwards, we can use the completion times of the last actions of each agent's plan to obtain an estimated execution time $\hat{T}_c$ of the whole plan before starting the execution.
Ideally, this should be equal to the cost of the plan as given by the MAPF solver.

However, $\hat{T}_c$ provides only the lower bound of the plan's \emph{real execution duration} $T_c$, as further delays during execution may increase it.
We can improve the estimate $\hat{T}_c$ during the execution as it progresses in real time by updating $G_{ADG}$.
When an agent completes an action $a$, we can replace the estimated action completion time $\hat{t}_c(a)$ with the \emph{real action completion time} $t_c(a)$.
Then, we can use the same procedure to update the estimated completion times of remaining actions.
After an agent $k$ finishes an action $a$, its corresponding vertex $V_{ADG}$ is updated with $t_c(a)$.
Then, we follow all outgoing edges to update $\hat{t}_s(a')$ and $\hat{t}_c(a')$ of all future dependent actions $a'$.

This way, the updated estimate is propagated not only to the rest of the agent's path via $E^1$, but thanks to $E^2$, also to all future actions of the agents that $k$ interacts with.
Following $E^1$ and $E^2$ also guarantees that only the dependent actions are updated.
By updating the completion times, we obtain a better estimate of the plan completion time $\hat{T}_c$ during execution.

\begin{figure}
    \centering
    \includegraphics[width=0.5\linewidth]{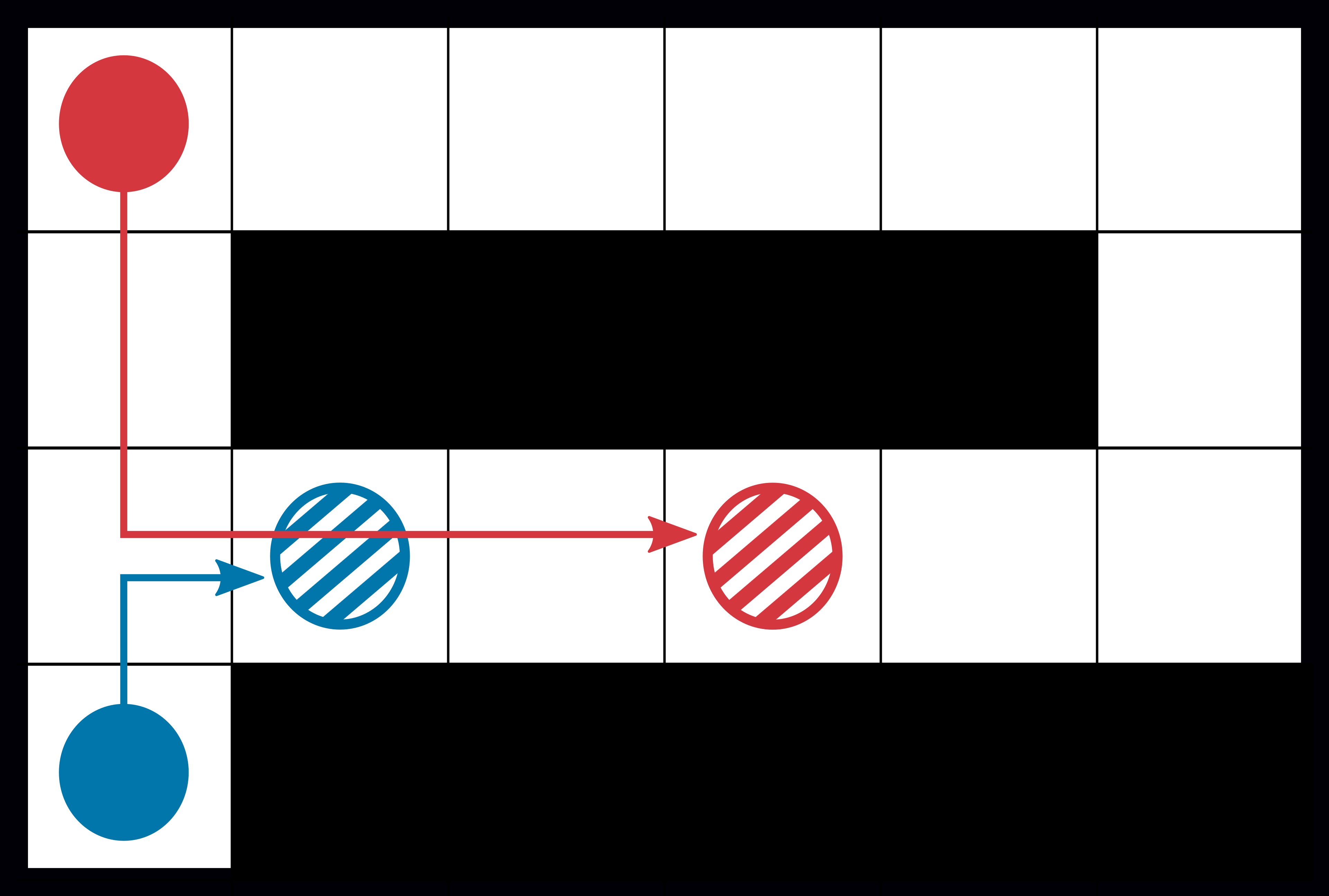}
    \caption{If only the red agent is significantly delayed, it should take the longer way around, allowing blue to reach its goal sooner. A modified version of example in~\cite{zahradka2023solving}.}
    \label{fig:replanning_example}
\end{figure}

\subsection{Resolving Delays}
As the execution progresses, each individual robot accumulates a delay at a different rate. 
If a part of the robotic fleet becomes delayed disproportionately to the rest of the fleet, the resulting configuration may not align with the original plan.
The accumulated delays may even become significant enough that there may now exist an alternate plan with lower execution cost $T_c$ than the one achieved by continuing the execution of the original plan.
This holds true even if the original plan is optimal.
We can obtain such a plan by \emph{rescheduling} or \emph{replanning}.

Consider the example in Fig.~\ref{fig:replanning_example}.
In the optimal plan, both robots drive through the middle corridor.
Because the corridor is narrow, the robots have to move in a single file.
If the blue robot drives first and reaches its goal, it would block the corridor for the red robot.
Therefore, the red robot drives first and blue waits for two time steps before following.
However, if the red robot gets delayed, blue has to wait longer than two time steps to prevent a collision.
This can be ensured using a robust execution method capable of retiming, but leads to an increased cost of the execution, because it causes the blue agent to accumulate delay too.
A similar case can be seen in the example solution in Fig.~\ref{fig:small_mapf_example_solution}.

If the delay of one of the robots is large enough, the execution cost may be reduced by either \emph{rescheduling} to switch the order of agents, or by \emph{replanning} to find new paths for the agents.
In Fig.~\ref{fig:small_mapf_example_solution}, rescheduling would be sufficient to allow the blue agent to cross earlier.
In fact, replanning would lead to the same outcome.
While it is often more costly to find a new set of paths instead of changing the ordering of the robots on the crossroads, replanning can be used to reschedule.
In the example in Fig.~\ref{fig:replanning_example}, however, rescheduling would not be able to lower the execution cost, because the blue agent cannot enter the corridor first.
Such case can be only solved by replanning to find a new path for the red agent.
Given a large enough delay of the red agent and sum of costs objective, $T_c$ can be decreased if the red agent takes the longer route through the upper corridor and the blue agent can move into its goal freely.

Having established that with robots facing individual delays during the execution, the fleet may enter a configuration where an alternate plan would lead to lower $T_c$, the problem is how to determine whether that applies for the current configuration of the fleet.
To know it for certain, we would need to have the alternate plan.
Since MAPF is an NP-hard problem~\cite{surynek2010optimization}, it is impractical to find new plans just to verify if they would be better than the current one.
It is therefore crucial to approximate or estimate the benefit that replanning would bring by other, simpler means.

\begin{figure}[htb]
    \centering
    \begin{subfigure}[b]{0.55\linewidth}
        \includegraphics[width=1.0\linewidth]{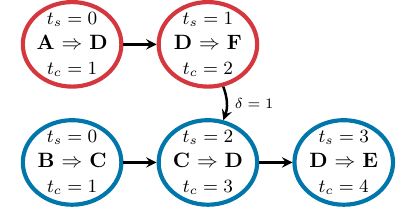}
        \caption{}
        \label{fig:adg_slack_edge}
    \end{subfigure}
    \begin{subfigure}[b]{0.35\linewidth}
        \bigskip
        \centering
        \includegraphics[width=1.0\linewidth]{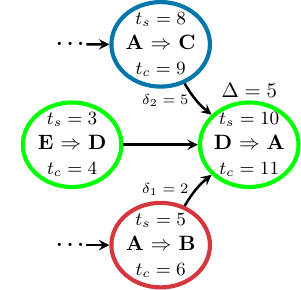}
        \caption{}
        \label{fig:adg_slack_vertex}
    \end{subfigure}
    \caption{(a) ADG with computed slack $\delta$ for a Type 2 edge. (b) Example of $\Delta$ computation for a robot (green) that moves into a vertex after two other agents (red and blue) pass through.}
    \label{fig:adg_slack}
\end{figure}

For this purpose, we suggest using the edges of $G_{ADG}$, which provide information about mutual dependencies of actions.
The Type 1 edges represent the dependencies between each agent's own actions and can be used to propagate information about the delay to the rest of the agent's path.
However, they cannot be used on their own to determine the potential benefit of replanning.
Assuming an optimal plan, replanning for a completely independent robot would not be beneficial even if it accumulates a significant delay.
That is because the robot is already on its optimal path and will arrive to its goal as soon as possible.
However, Type 2 edges contain information about interactions between different robots.
If the delayed robot's actions have any outgoing Type 2 edges, they can be used to express how much is the delayed robot affecting other members of the fleet by progressing throughout the plan slower.
Rescheduling or replanning may be beneficial in this case, because instead of waiting for the delayed robot, the others could be rescheduled in front of it or find another path to their goal.

For every Type 2 edge $E^2 = (a_i^{r_1}, a_j^{r_2})$ between the actions of robots $r_1$ and $r_2$ we can calculate a \emph{time reserve}, or \emph{slack}, $\delta(E^2) = t_c(a_i^{r_1}) - t_c(a_{j-1}^{r_2})$.
An example of computed $\delta(E^2)$ can be seen in Fig.~\ref{fig:adg_slack_edge}.
Slack can be computed during the completion time estimation procedure, directly after line 12 of Algorithm~\ref{alg:time_predictions}.
It is a simple way to estimate how long $r_2$ must wait before it can start an action due to the temporal dependence on $r_1$.
The $\delta(E^2)$ is interpreted as positive for the vertices on the head end of $E^2$ (actions that must wait) and negative for the vertices on the tail end (actions that cause waiting).
In other words, when $r_2$ waits for $r_1$ to finish an action, the action of $r_2$ would have a positive slack on an incoming $E^2$, while the action of $r_1$ would have a negative one on an outgoing $E^2$.
On the other hand, if $r_2$ was originally planned to move later than $r_1$ and therefore does not have to wait, $r_2$ would have negative slack on the incoming edge, and $r_1$ would have positive slack on the outgoing edge.
In such a case, $\delta$ expresses how much $r_1$ can be delayed without affecting $r_2$.
Such situations are not interesting for replanning, and therefore we refer to positive slack only to cases where the positive slack is on an incoming edge and negative slack only when it is on an outgoing edge.

Each $V_{ADG}$ may have multiple incoming $E^2$, which means that there is temporal precedence of multiple actions of potentially many agents in the same location.
Since $\delta(E^2)$ can only capture dependencies between two agents, we define the \emph{slack of an action} as $\Delta(a) = \max\{\delta(E^2): E^2 = (a', a) \in E_{ADG}\}$.
The $\Delta(a)$ expresses how long the agent will have to wait before it can start executing $a$.
An example can be seen in Fig.~\ref{fig:adg_slack_vertex}.

As the execution progresses, the estimates $\hat{t}_c(a) : a \in V_{ADG}$ are updated with the real measured action completion times $t_c(a)$.
Consequently, $\delta(E^2)$ and $\Delta(a)$ are also updated to reflect the current situation and provide an estimate of the necessary waiting times in the rest of the plan.
However, the expected waiting times themselves are not sufficient to decide whether an intervention could be beneficial.
Optimal plans often contain waiting actions, as seen in examples in Figs.~\ref{fig:small_mapf_example} and~\ref{fig:replanning_example}.
In such cases, either an alternate path for the agent would be longer than the necessary wait, or it may not even exist.
Therefore, we want to compare the current action slack with the \emph{initial action slack} $\Delta_0$, calculated at the start of the execution, to obtain $\bar{\Delta}(a) = \Delta(a) - \Delta_0(a)$.
The value of $\bar{\Delta}(a)$ tells us whether an agent will have to wait longer than originally planned due to delays of other agents. 
This computation differs from the original presented in~\cite{zahradka2025holistic}, where we computed the initial slack for each $E^2$, producing $\delta_0$, and used these values to compute the action slack $\Delta$.
The original equation took into account any agent that is significantly delayed and can influence other agents, but not necessarily does.
The improved version presented in this paper takes into account only the agents that actually do impact another agent.

The state of the whole execution considering all agents can then be expressed as $\bar{\Delta}^F = \text{max}_{a \in V}(\Delta(a))$.
A positive value of $\bar{\Delta}^F$ means that there is an agent that will have to wait longer than originally planned to execute an action because of a delay of another agent.
If the value is large, it suggests that the deviation from the original plan is significant and that it impacts some agent.
Therefore, there is a possibility that there is an alternate plan which could lead to a lower execution cost.
To decide whether we should try replanning or not, we can set a threshold value $\bar{\Delta}^T$, and when $\bar{\Delta}^F > \bar{\Delta}^T$, rescheduling or replanning can be initiated.

Thus, using ADG with slack-augmented vertices, we obtain a \emph{holistic} architecture to execution optimization.
The execution is safe and robust due to retiming performed by ADG.
As the execution progresses, we can update the expected action and plan completion times by propagating updates via Type 1 and Type 2 edges of the ADG. 
If some robot is delayed enough in a way that affects the rest of the fleet, as measured by $\bar{\Delta}^F$, we can attempt to reduce the execution time by rescheduling or replanning.
The frequency of rescheduling and replanning can be controlled by the threshold value, which determines what delays are too significant to be solved by retiming.
The decision whether to replan or reschedule can be made by first using the cheaper alternative of rescheduling, and if it produces the same plan as the original one, we can try the more costly replanning.

\section{Experimental Environment}
In this section, we describe the tools that we used to experimentally verify the developed architecture.
First, we describe our real-life demonstrator on which we verify that the robust execution monitoring and optimization architecture is capable of safe operation of a robotic fleet.
Then, we describe the simulator we developed to conduct large-scale experiments for verification of the method's optimization performance.
Finally, we describe the process we use to model delays.

\subsection{Real-life Demonstrator}

\begin{figure}
\centering
    \begin{subfigure}[t]{0.49\textwidth}
        \centering
        \includegraphics[width=1.0\textwidth]{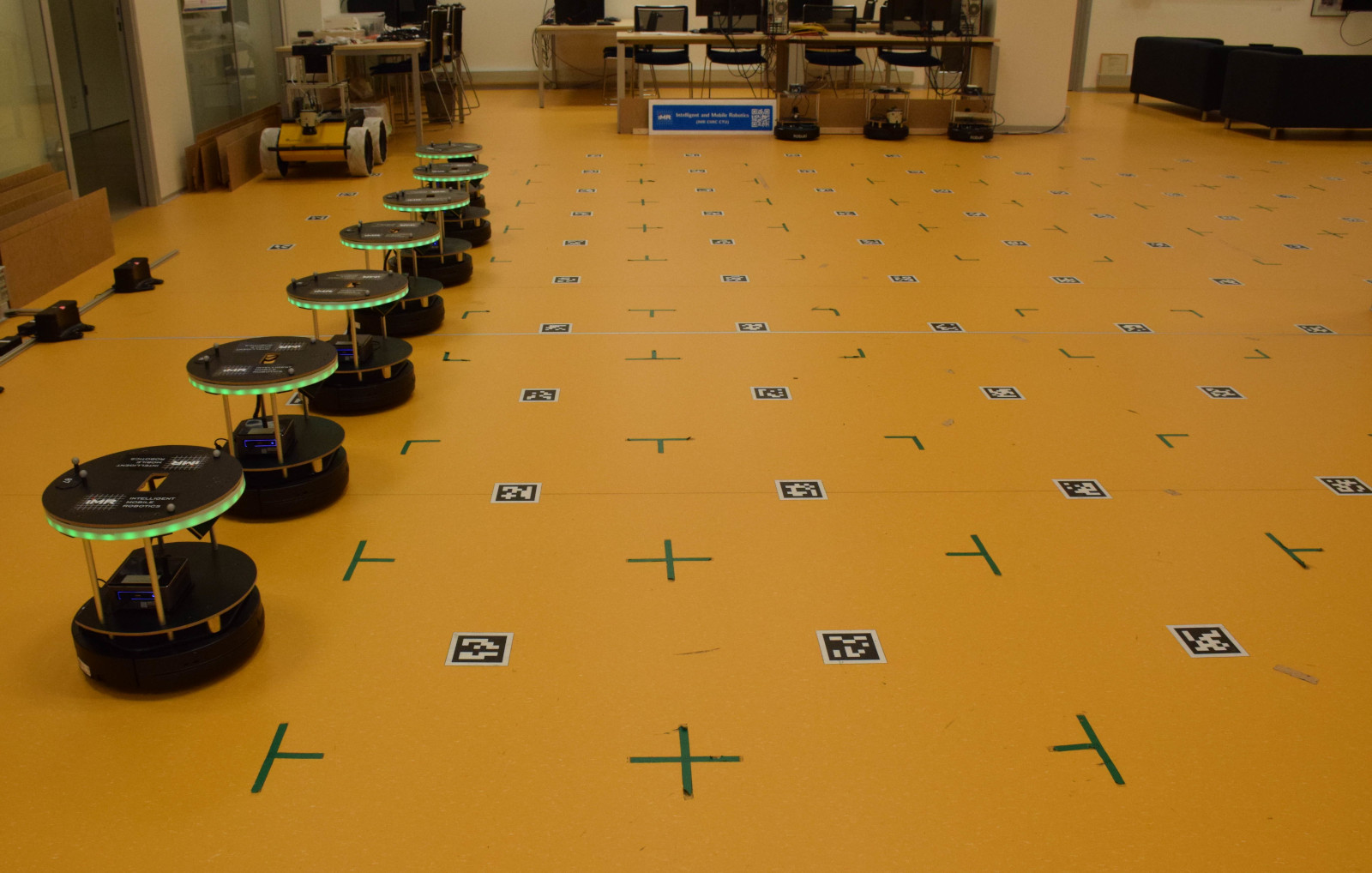}
        \caption{A robotic fleet ready to start execution.}
        \label{fig:demonstrator_ready}
    \end{subfigure}
    \begin{subfigure}[t]{0.49\textwidth}
        \centering
        \includegraphics[width=1.0\textwidth]{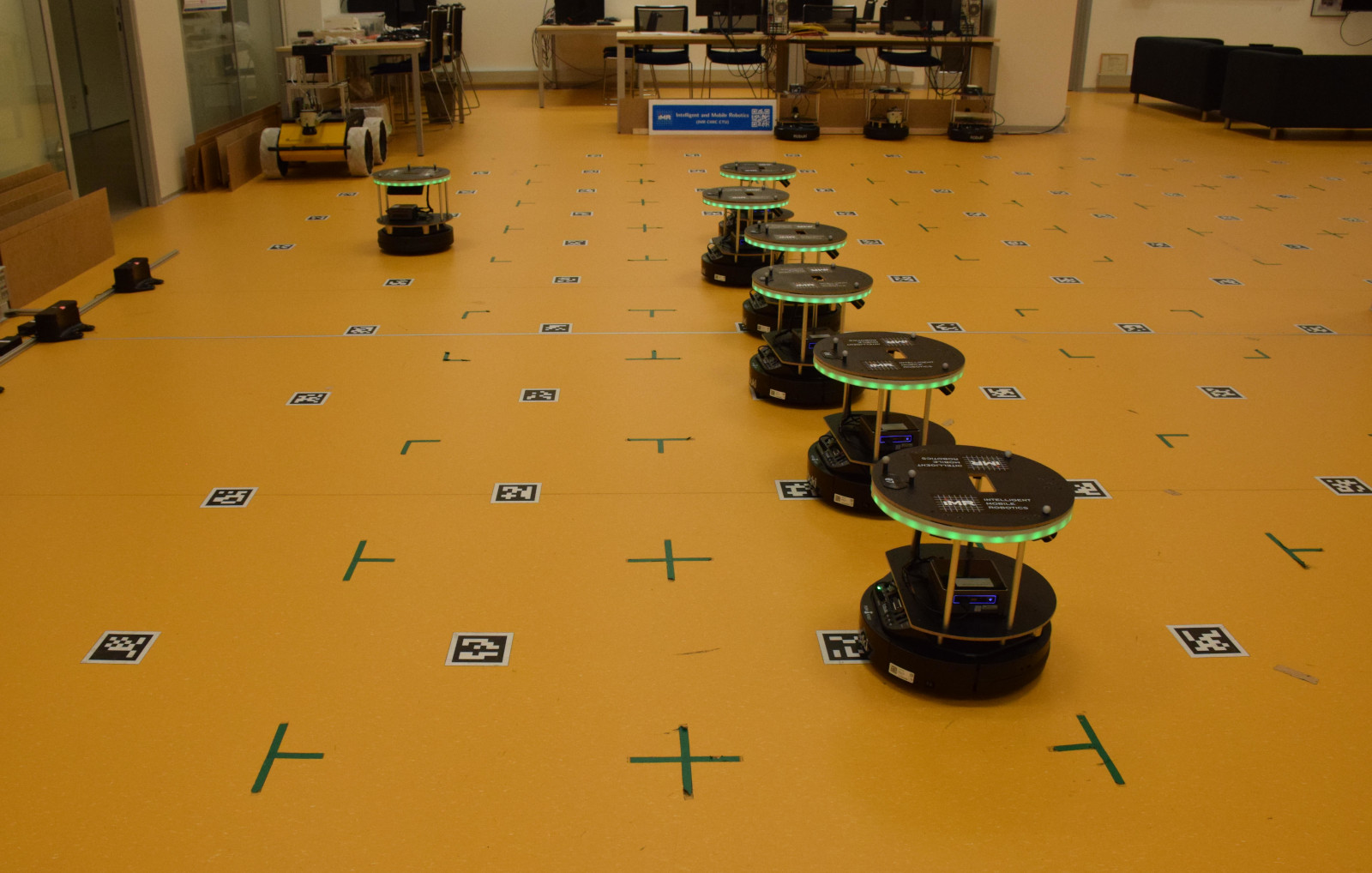}
        \caption{Execution in progress.}
        \label{fig:demonstrator_driving}
    \end{subfigure}
    \begin{subfigure}[t]{0.49\textwidth}
        \centering
        \includegraphics[width=1.0\textwidth]{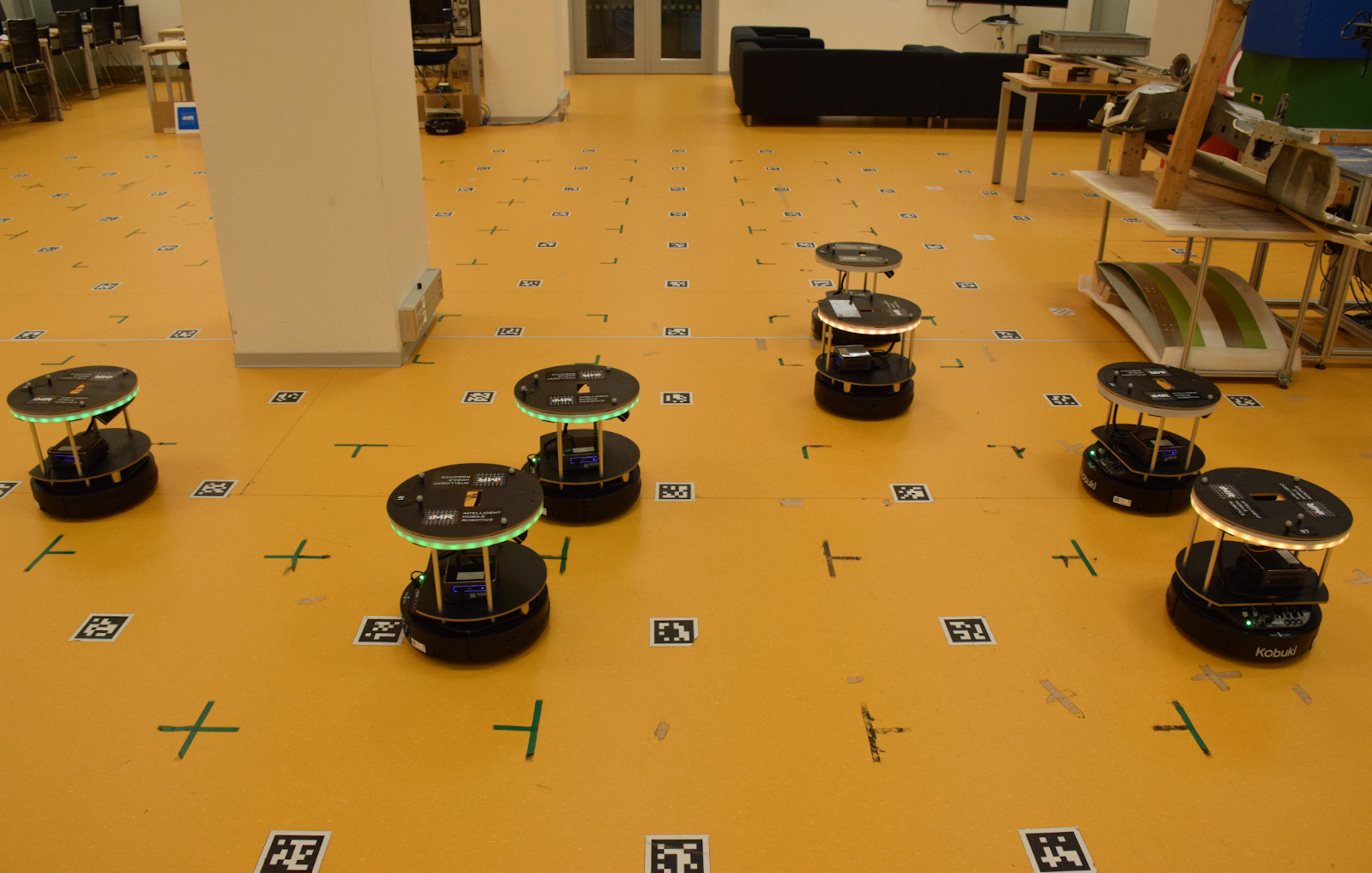}
        \caption{Area with high congestion.}
        \label{fig:demonstrator_congestion}
    \end{subfigure}
    \begin{subfigure}[t]{0.49\textwidth}
        \centering
        \includegraphics[width=0.85\textwidth]{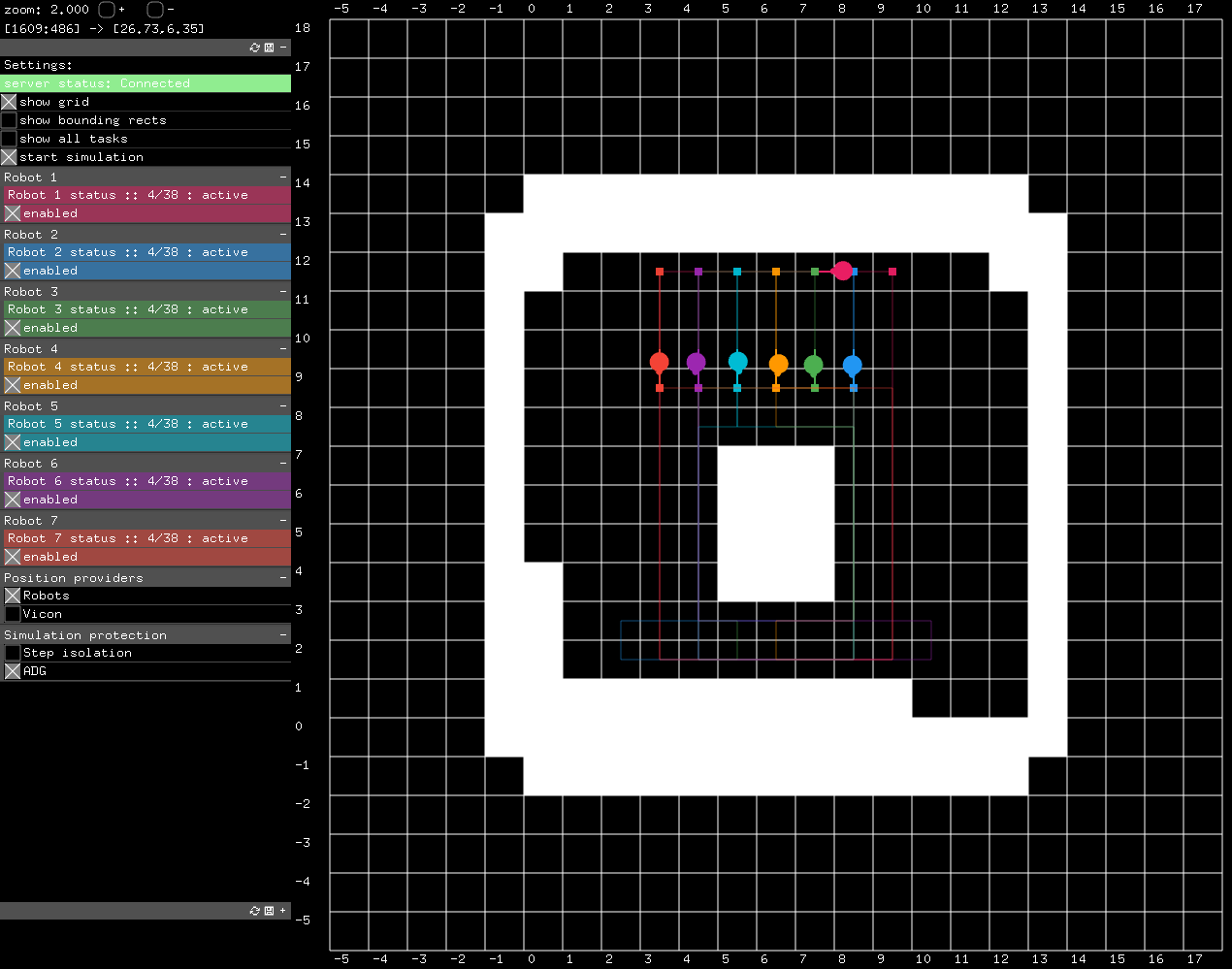}
        \caption{GUI. The robotic fleet is in the configuration seen in (b).}
        \label{fig:demonstrator_interface}
    \end{subfigure}
    
    \caption{Robotic fleet demonstrator developed for real-life verification of execution optimization methods.}
    \label{fig:demonstrator_basic}
\end{figure}

To verify how MAPF and robust execution methods work in practice, we have built a real-life demonstrator of an autonomous warehouse robotic fleet shown in Fig.~\ref{fig:demonstrator_basic}.
It consists of $7$ autonomous robots based on the TurtleBot2 that can be seen in Fig.~\ref{fig:demonstrator_ready} and a centralized server, whose role can be filled by any standard computer or laptop.

Each robot has the responsibility to localize itself, execute actions given by the server and report back their completion.
The actions are the same actions that are used in ADG.
More concretely, each action contains a starting position and orientation and a goal position and orientation.
The server's responsibility is to maintain a robust execution method such as ADG, send available actions to the robots, receive reports about their completion, and update the robust execution method accordingly.
The responsibilities are strictly divided, and each component has access only to the information it requires.
Only the server has the whole plan for each robot, and the robots do not perceive or communicate with each other at all.
On the other hand, the server does not have information about the location of each robot.

There is an optional third element in the form of a GUI, which is used to launch and pause the execution, display the configuration of the fleet and its status and adjust parameters of the execution, such as which execution method is being used.
The GUI can be seen in Fig.~\ref{fig:demonstrator_interface}.
To visualize the fleet configuration, each robot must be broadcasting its current position.
However, the GUI is the only part of the system that collects this information. 

The robots localize themselves using odometry combined with a monocular RGB camera, which scans AprilTags~\cite{wang2016apriltaga} located in a grid pattern on the ground.
The AprilTag grid (white squares with black patterns) can be seen in front of the robots in Fig.~\ref{fig:demonstrator_ready}.
Each robot is also equipped with a bumper, which is used to detect if there is an obstacle in the way.
If the bumper hits an object, the robot pauses the execution of the current command for $5$ seconds before continuing again.

The main loop of the server is straightforward.
Let us assume that we are using ADG as a robust execution method.
In each iteration, the server checks whether there is any agent that is not executing an action and has some actions remaining in its plan.
If there is such an agent, the server checks whether it can start executing its next action.
That means checking whether all the actions associated with the next action's incoming edges are completed.
If that is the case, the server sends the next action to the agent and labels the agent as \texttt{executing}.
Then the server continues with the next agent.

When an agent completes an action, it sends a confirmation report back to the server.
The server handles the reception of the message with a callback that marks the corresponding action in ADG and its outgoing edges as completed and removes the \texttt{executing} label from the agent.

\subsection{Simulator}
Because the real-life demonstrator does not allow for conducting large-scale experiments, we developed a real-time simulator with the same architecture used by the demonstrator.
The main loop of the simulator is the same: the program iterates through idle agents to check whether one of them has an action in its plan that it can start executing.
This means checking whether there is an action that has all precedence constraints fulfilled.
If there is such an action, the main loop spawns a thread which simulates the execution of the action (instead of sending a message to a robot), marks the agent as busy and continues with the next idle agent.

The thread that simulates the action's execution first checks whether the goal vertex of the action is free or if there is something currently blocking it.
If the goal vertex is occupied, the thread waits $\SI{100}{\ms}$ before checking again.
Otherwise, the execution continues.
To remove any influence of the dynamics of the simulated robots on the execution cost, we assume perfect execution of actions.
This means that all possible actions -- that is, movement and rotation by $\ang{90}$ -- take exactly $\SI{1}{\s}$.
After $\SI{1}{\s}$ elapses, the thread reports to the ADG that the action was completed before marking the agent as idle and exiting.
Since the ADG is shared between many threads, access is protected by a mutex.

\begin{figure}
    \centering
    \includegraphics[width=0.4\linewidth]{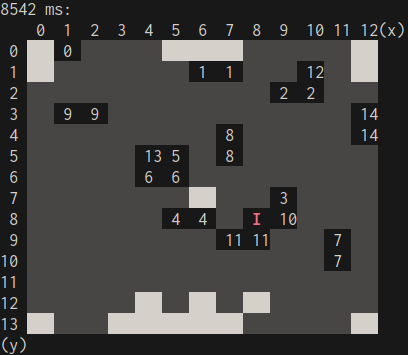}
    \caption{An optional plan execution visualization provided by the simulator. White areas are obstacles and grey areas are free space. Cells with numbers represent agents. An agent is shown to occupy two cells whenever it is moving between them. Red I shows intruder blocking the agent 10.}
    \label{fig:simulator_interface}
\end{figure}

The decision whether to replan or not is made by another thread, which monitors the state of the execution. 
When replanning is triggered, it uses a boolean flag to inform the other currently running threads.
The main loop will stop assigning new actions, and once every agent completes its last assigned action, a new plan is found starting from the latest configuration of agents.
If an agent is executing an action and is not blocked, we wait until the agent completes the action and use the action's goal location.
We use the agent's current location otherwise.
A new ADG is created from the new plan, which is used once the execution continues.
SOC is measured by summing the completion times of the last actions of the agents, and makespan can be measured by taking the maximum completion time among all agents. 
In case of replanning, the time the other robots had to wait for the last robot to stop is also included in SOC.
The maximum difference between the SOC and the makespan of the original plans and the values measured by the simulation was less than $0.5\%$, with the simulated values being slightly higher.
The simulator also features an optional visualization, which can be seen in Fig.~\ref{fig:simulator_interface}.

\subsection{Modelling Delays}
\begin{figure}
\centering
    \begin{subfigure}[t]{0.49\textwidth}
        \centering
        \includegraphics[width=1.0\textwidth]{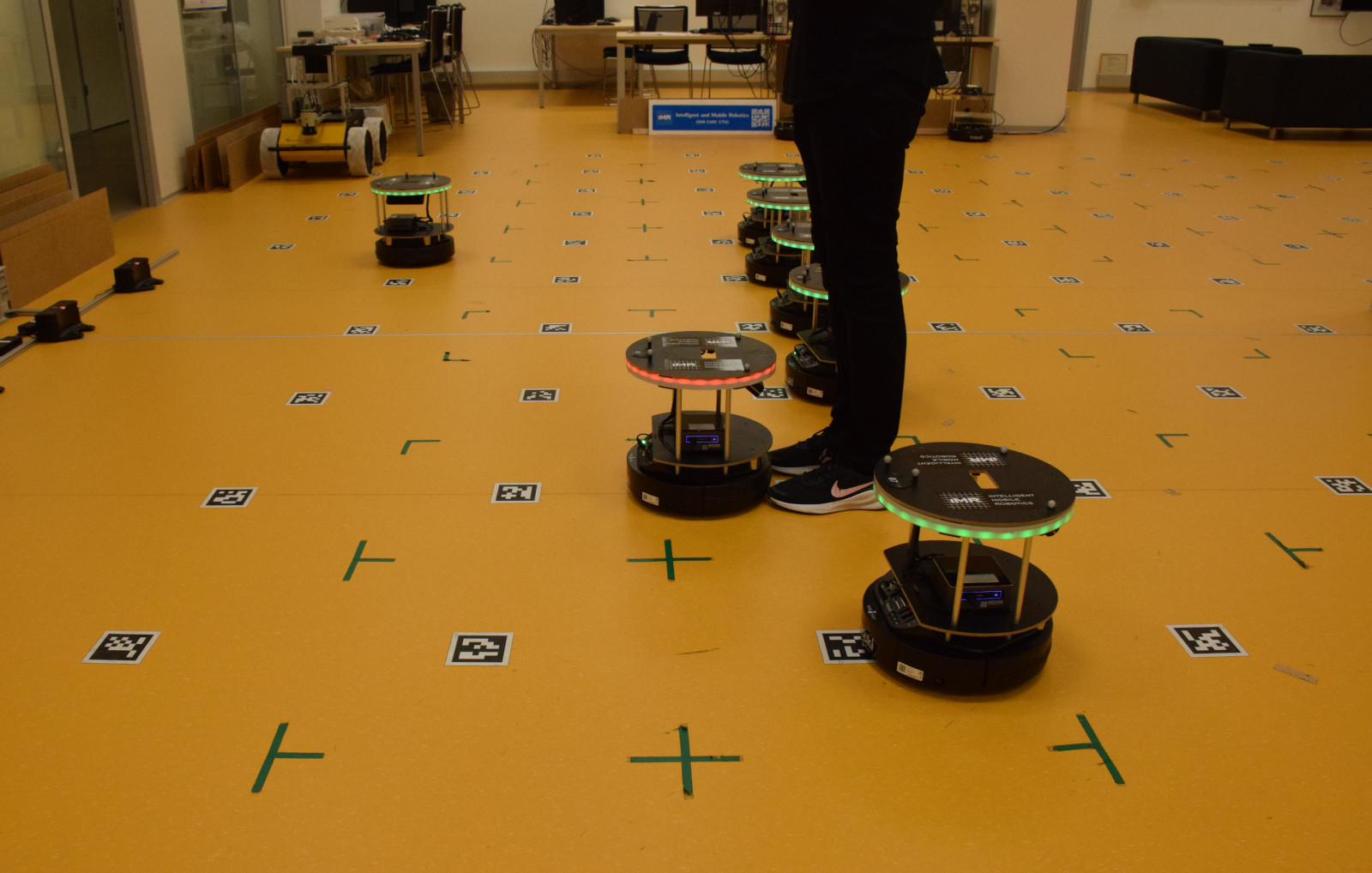}
        \caption{A robot being blocked by an intruder. Indicated by red light.}
        \label{fig:demonstrator_intruder}
    \end{subfigure}
    \begin{subfigure}[t]{0.49\textwidth}
        \centering
        \includegraphics[width=1.0\textwidth]{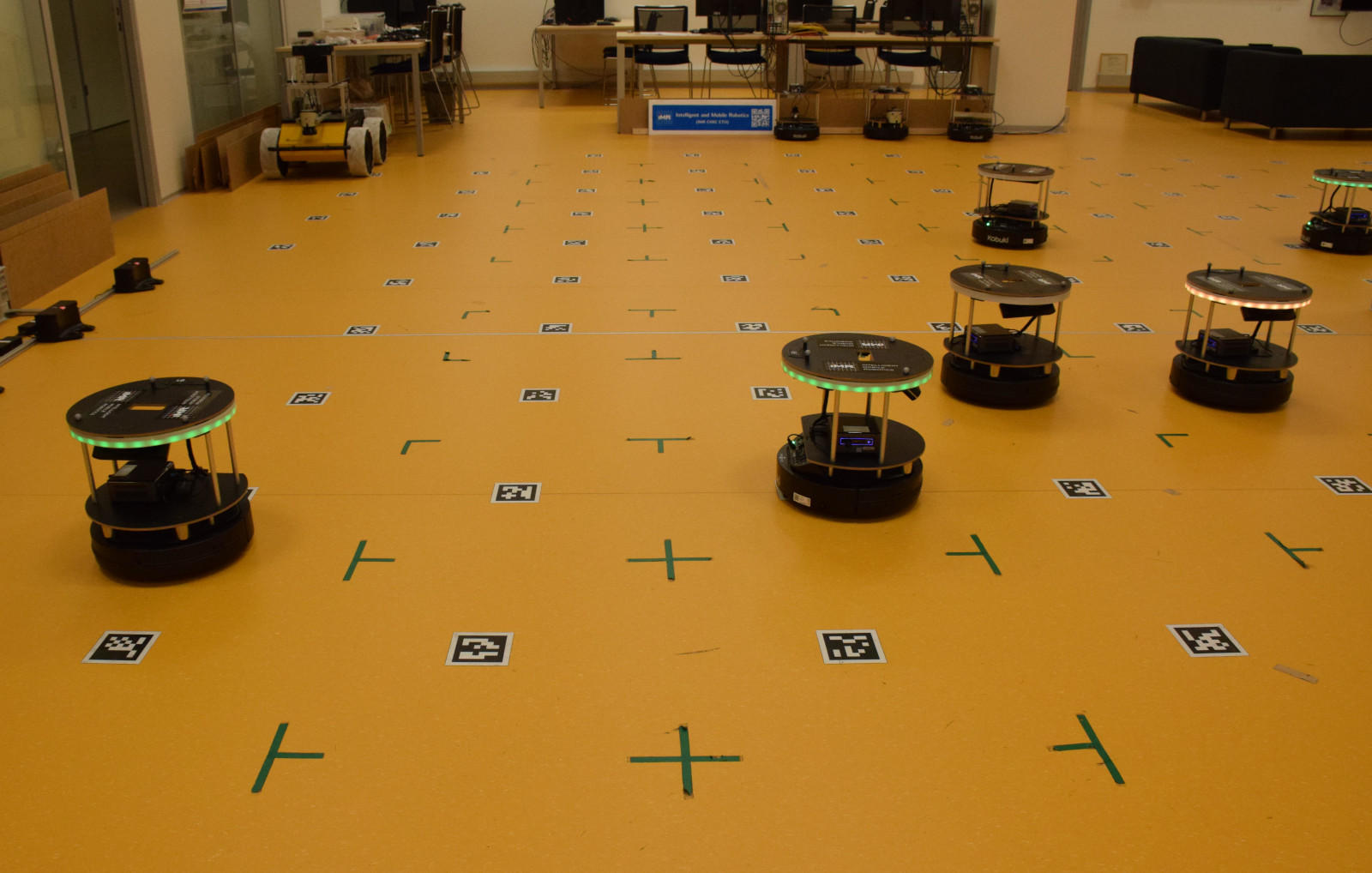}
        \caption{Desynchronized fleet because of intruder's interference. A robot is waiting for the delayed robot as indicated by the shining orange light (middle right).}
        \label{fig:demonstrator_desync}
    \end{subfigure}
    \caption{Experiments with an intruder.}
    \label{fig:demonstrator_intruder_experiments}
\end{figure}

In order to introduce a delay to an agent, we use the so-called \emph{intruder}.
An intruder is an unplanned agent that enters the environment and occupies a vertex, making it untraversable for the agents.
We assume that the intruder enters a specific vertex for a fixed amount of time before leaving.
We choose not to model the intruder's path in and out of the vertex because we are interested only in the situations where the intruder blocks an agent.
However, it is possible to either find a path for the intruder from some appearance location to some disappearance location, blocking different vertices at different times, or even use a random walking intruder.

An example of a real-life experiment with an intruder can be seen in Fig.~\ref{fig:demonstrator_intruder_experiments}.
A human enters the agents' operational space, e.g., for a maintenance task, and blocks a vertex that an agent is supposed to go through.
The agent registers that it cannot enter the vertex and waits until the intruder leaves before continuing with its plan, or until a new plan is found by replanning and the agent receives a new action with a different goal.

We replicate the same behavior in the simulator.
The intruder has two parameters: appearance time $T_a^I$ and disappearance time $T_d^I$.
Then, it selects a vertex on the path of a random agent that is not yet in its goal, such that the agent is scheduled to visit the vertex in $T_a^I + \SI{2}{\s}$.
The two-second offset ensures that the intruder has enough time to enter before the selected agent even if some other agents are moving through the vertex.
The intruder disappears again after $T_d^I$ elapses.
If replanning occurs while the intruder is active, the intruder continues occupying the vertex even after the replan until its scheduled disappearance.
We assume that neither $T_a^I$, $T_d^I$ nor the intruder's position are known to the system.

\section{Experimental Evaluation}

\begin{figure}[htb]
\centering
    \begin{subfigure}[c]{0.2225\textwidth}
        \centering
        \includegraphics[height=2.9cm]{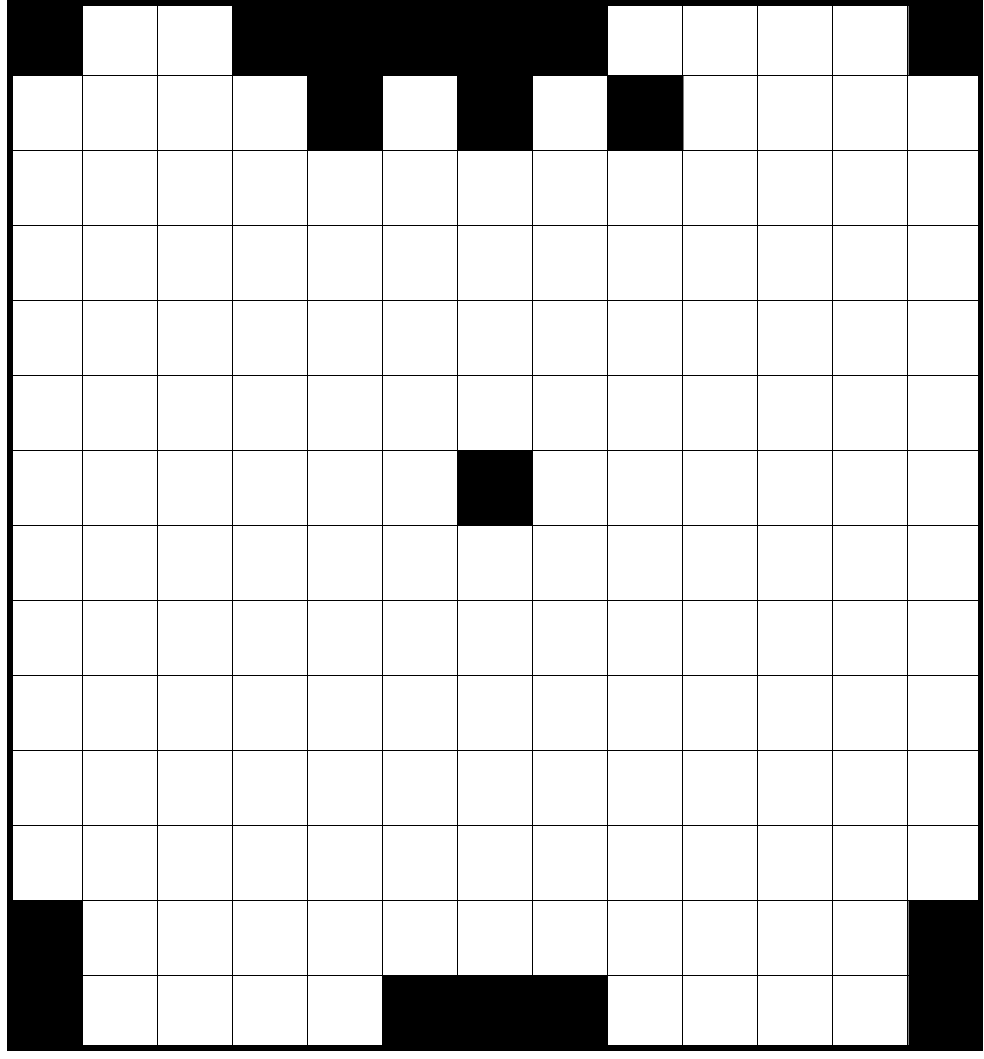}
        \caption{\texttt{lab}}
    \end{subfigure}
    \begin{subfigure}[c]{0.24\textwidth}
        \centering
        \includegraphics[height=2.9cm]{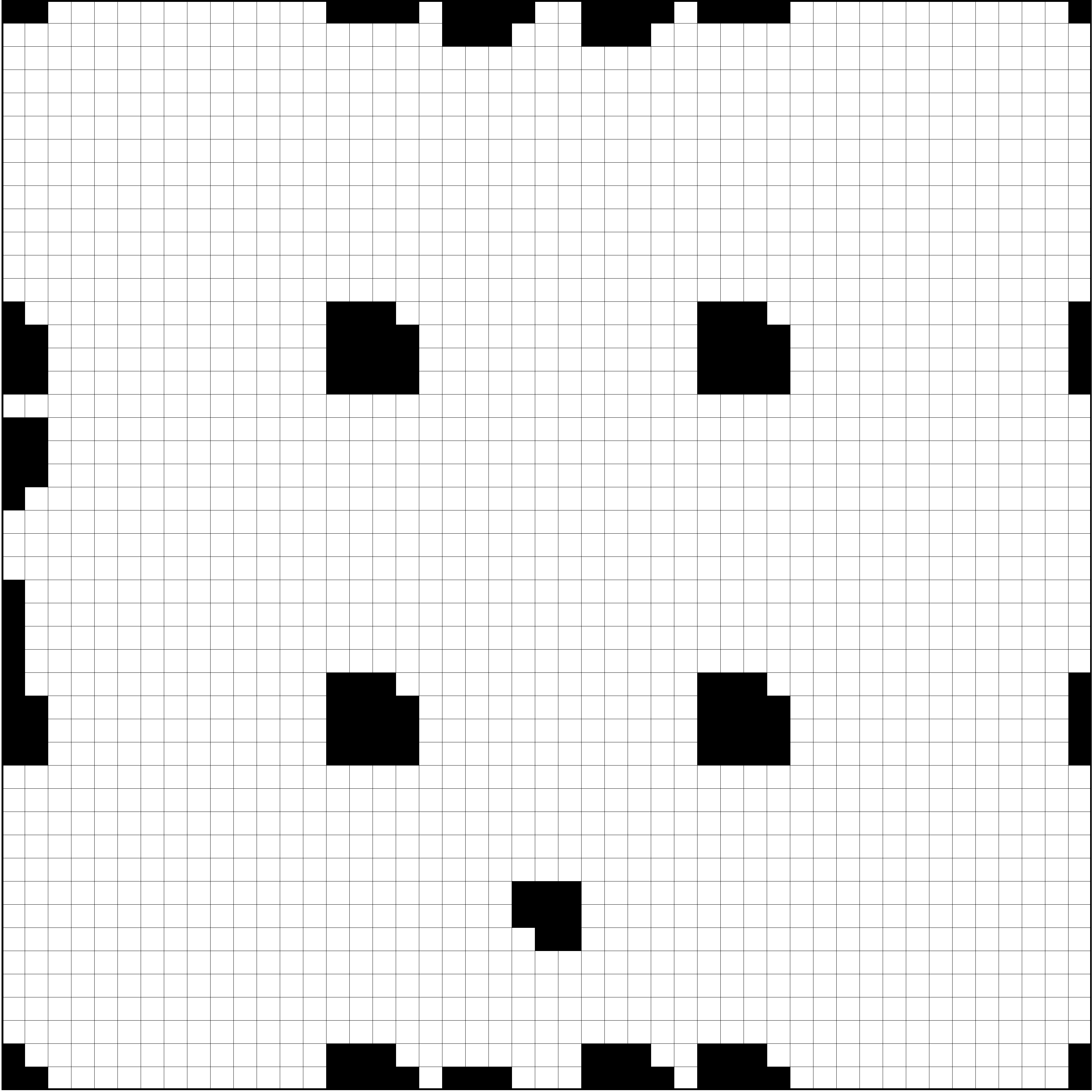}
        \caption{\texttt{arena}}
    \end{subfigure}
    \begin{subfigure}[c]{0.24\textwidth}
        \centering
        \includegraphics[height=2.9cm]{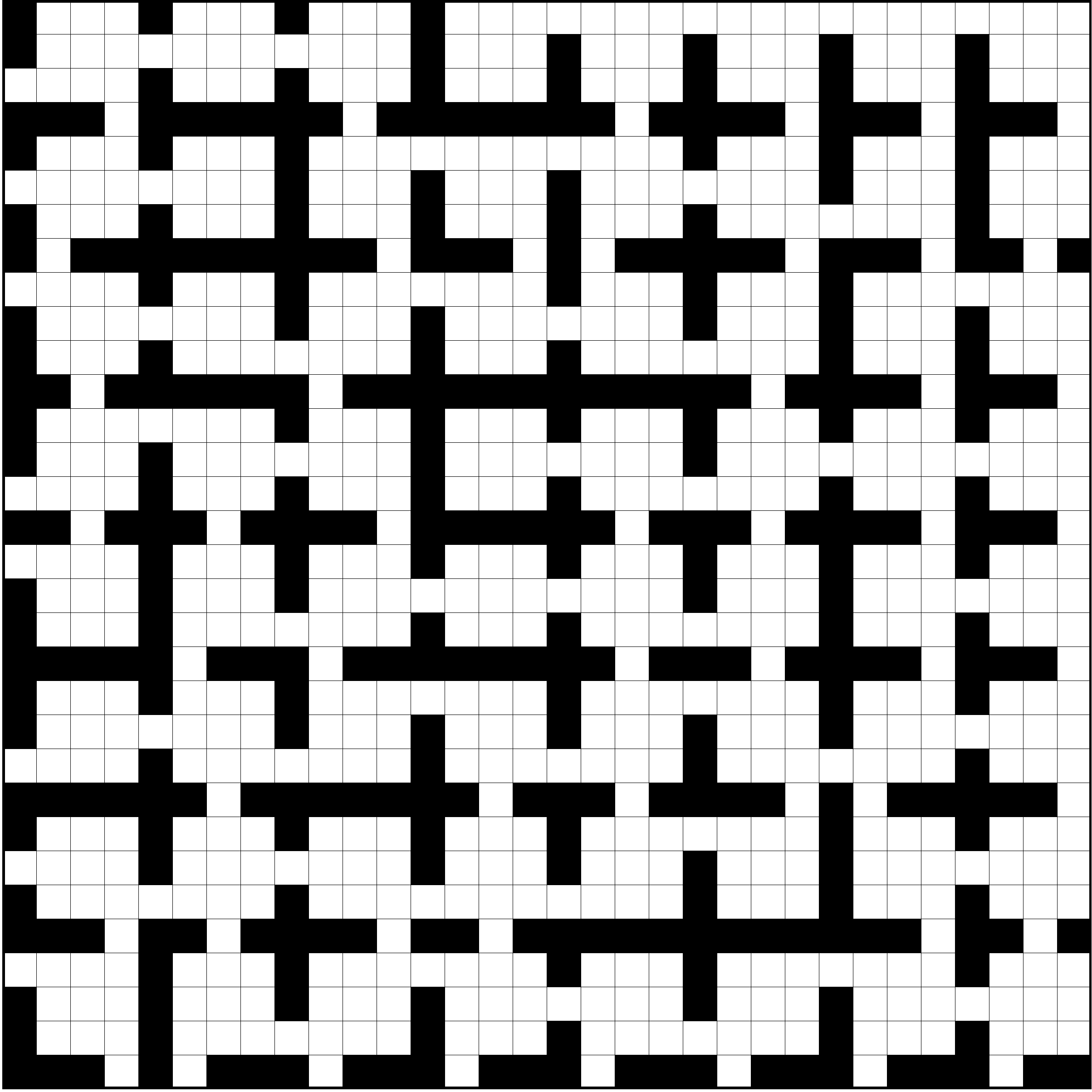}
        \caption{\texttt{room32}}
    \end{subfigure}
    \begin{subfigure}[c]{0.24\textwidth}
        \centering
        \includegraphics[height=2.9cm]{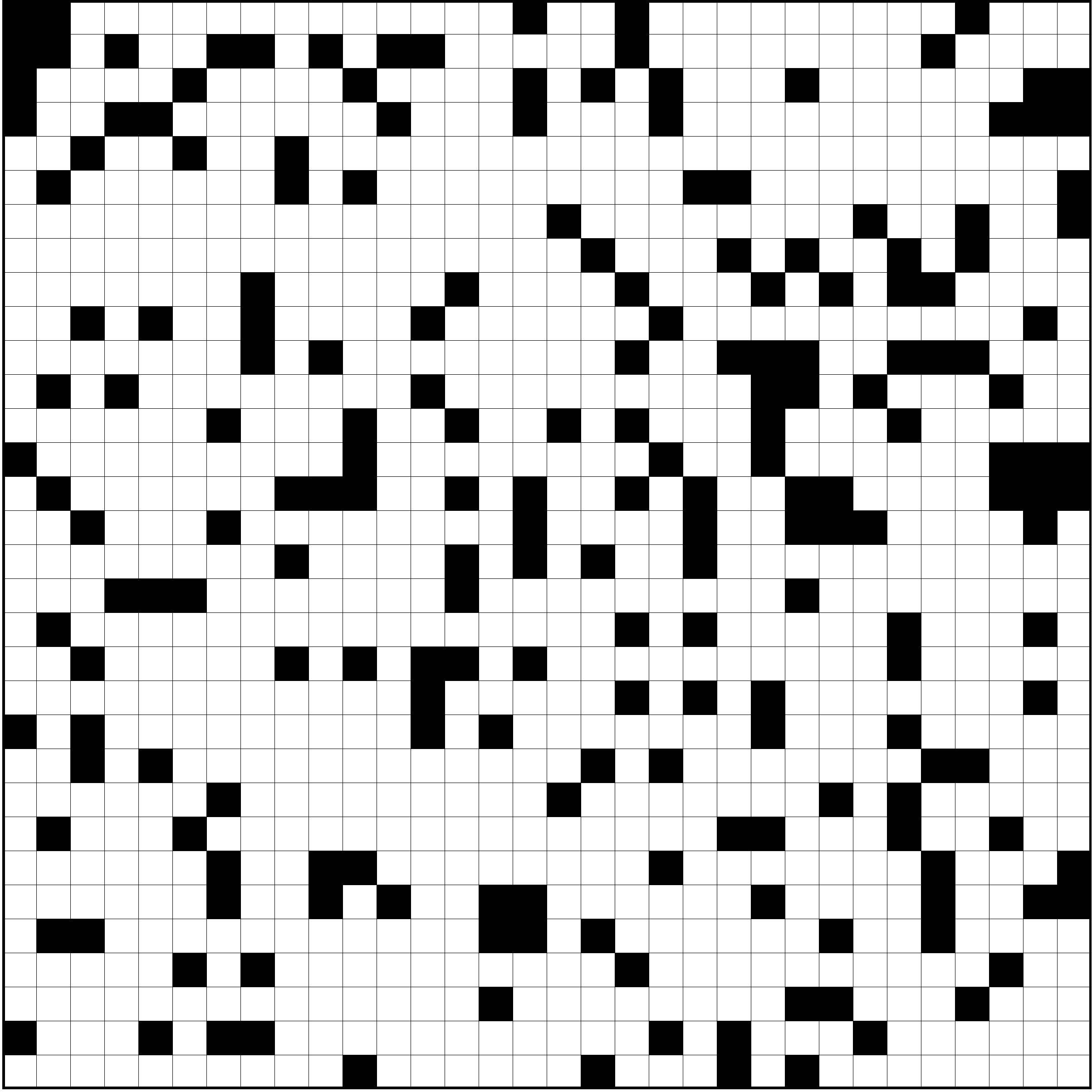}
        \caption{\texttt{random32}}
    \end{subfigure}
\caption{Maps used in our experiments. Black cells are obstacles, white cells are free to traverse.}
\label{fig:example_maps}
\end{figure}

\subsection{Instances}
We used four maps shown in Fig.~\ref{fig:example_maps}.
The first map (\texttt{lab}) represents our laboratory.
The second map (\texttt{arena}) is a large, mostly freely traversable map\footnote{\texttt{arena} map sourced from \url{https://github.com/Kei18/mapf-IR}.}.
The last two maps (\texttt{room32-32-20} and \texttt{random32-32-20}) are from the widely used MovingAI MAPF benchmark dataset~\cite{stern2019multi}.
They will be refered to as \texttt{room32} and \texttt{random32}, respectively.

Next, for each map, we generated random instances with varying numbers of agents such that they can be reliably solved with a 1-robust CBS solver.
On \texttt{lab}, \texttt{room32} and \texttt{random32}, we used $5$, $10$ and $15$ agents, and on the more open \texttt{arena} map, we used $10$, $15$, $20$ and $25$ agents.
Then, we generated $20$ different instances for each map and number of agents, producing 60 instances for \texttt{lab} and \texttt{room32} each, and $80$ instances for \texttt{arena}.
Each instance was tested with $5$ different randomization seeds.
In total, we ran $1300$ experiments.

To obtain plans for each map, we used a 1-robust modification of CBS that produced SOC-optimal solutions.
The same solver was used for replanning. 
We did not use a separate rescheduling procedure and used only replanning.
However, since replanning is a more general problem than rescheduling, a planning solver can also reschedule when it finds the same set of paths, but with different timing. 

Each experiment consisted of four parts: i) simulation without an intruder to obtain the lower bound of execution cost $T_c^{lb}$; ii) simulation with an intruder; iii) simulation with an intruder and replanning at a random time; and iv) simulation with an intruder and Slack-based predictive replanning.
SOC was measured in milliseconds.
The random replanning baseline method selected a random time to replan between the time the intruder appears $T_s^I$ and $T_c^{lb} - \SI{3}{\s}$. 
This is because we assume that replanning before the intruder appears or too late into the execution is almost guaranteed to achieve no notable improvement.
Replanning with the Slack-based predictive approach occured as soon as the maximum slack $\bar{\Delta}^F$ crossed the threshold $\bar{\Delta}^T=\SI{2000}{\ms}$.
Both methods could replan at most once in each experiment.
The random baseline method was guaranteed to replan in every experiment, the predictive approach could decide not to replan and continue with the original plan.

The intruder was set to appear at time $T_{s}^I = \SI{3000}{\ms}$ and disappear at time $T_{e}^I=\SI{10000}{\ms}$, blocking one agent.
The blocked agent was a randomly selected agent that is not yet in its goal.

\subsection{Replanning Decision Efectiveness}

\begin{figure}[htb]
\centering
    \begin{subfigure}[b]{0.49\textwidth}
        \centering
        \includegraphics[height=3.9cm]{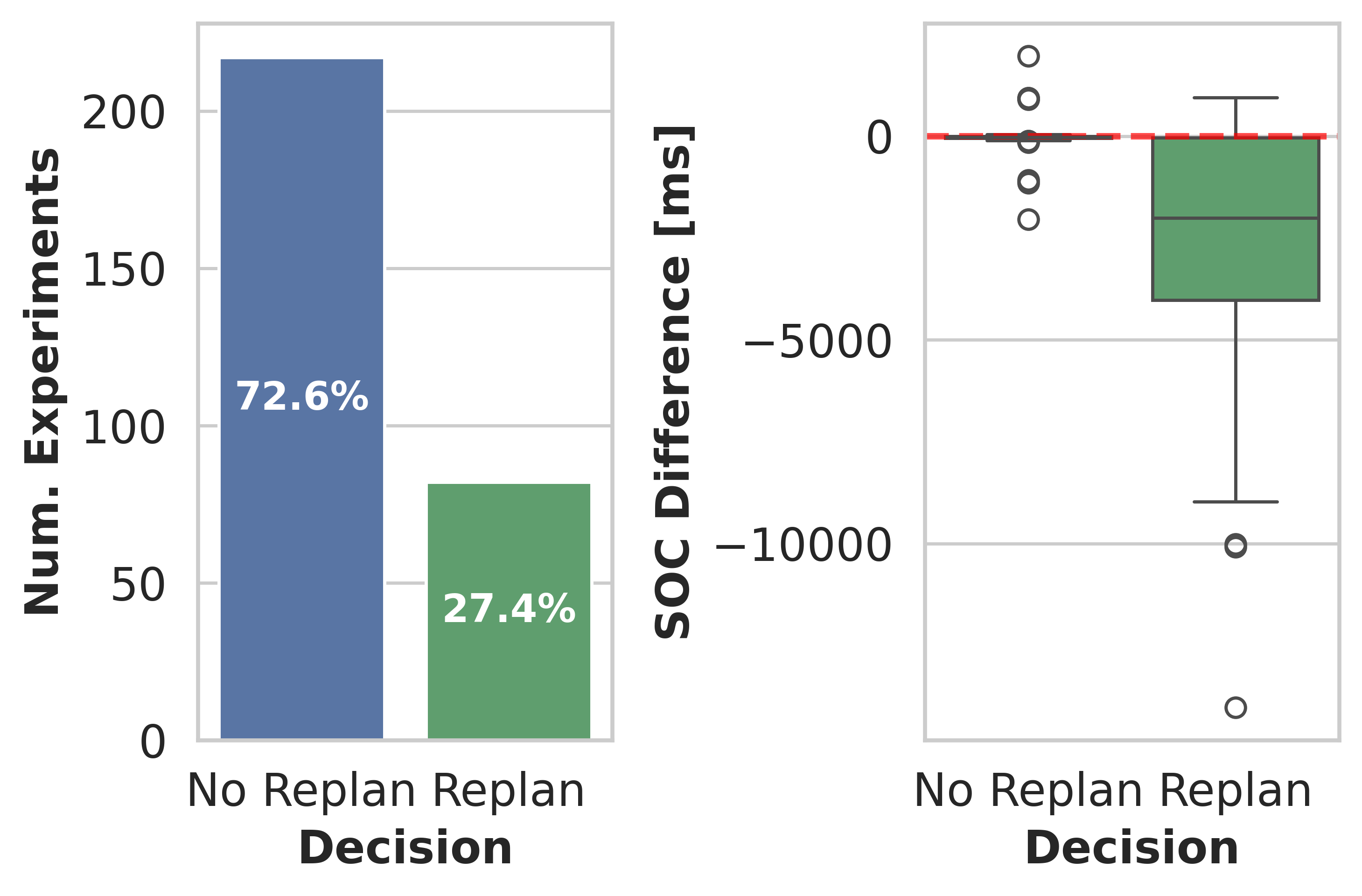}
        \caption{\texttt{lab}.}
        \label{subfig::impact_lab}
    \end{subfigure}
    \begin{subfigure}[b]{0.49\textwidth}
        \centering
        \includegraphics[height=3.9cm]{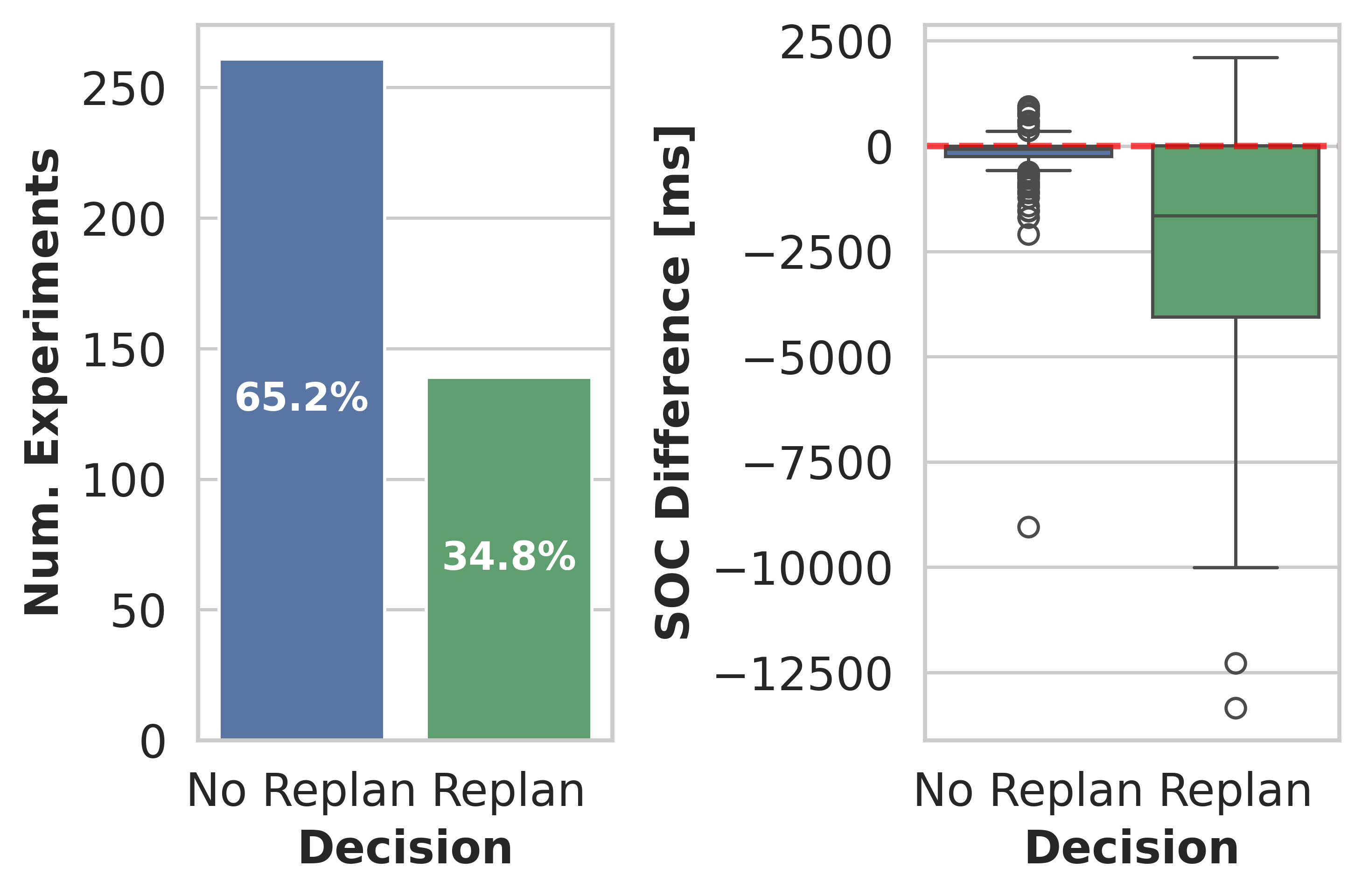}
        \caption{\texttt{arena}.}
        \label{subfig::impact_arena}
    \end{subfigure}
    \begin{subfigure}[b]{0.49\textwidth}
        \centering
        \includegraphics[height=3.9cm]{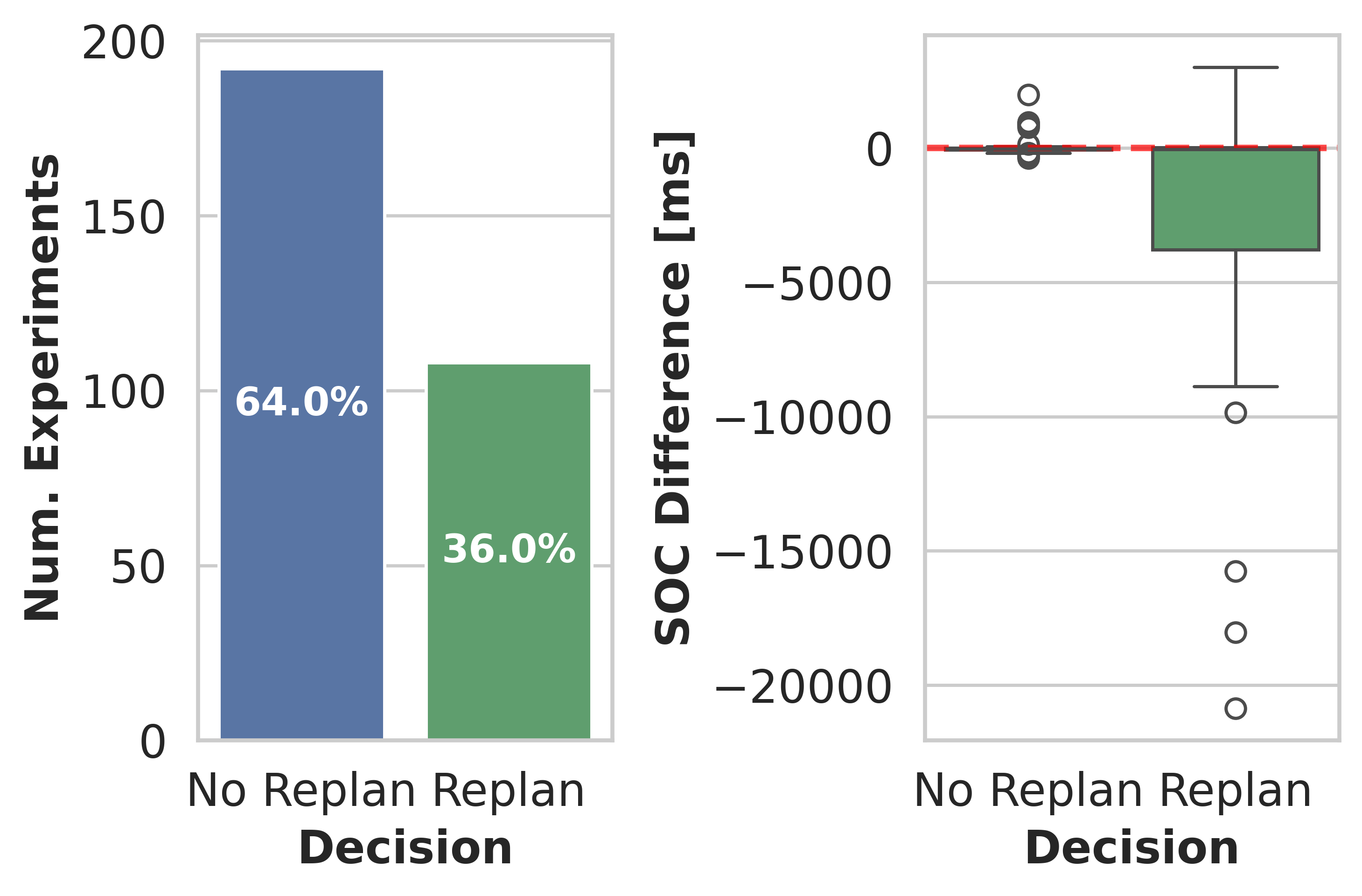}
        \caption{\texttt{room32}.}
        \label{subfig::impact_room32}
    \end{subfigure}
    \begin{subfigure}[b]{0.49\textwidth}
        \centering
        \includegraphics[height=3.9cm]{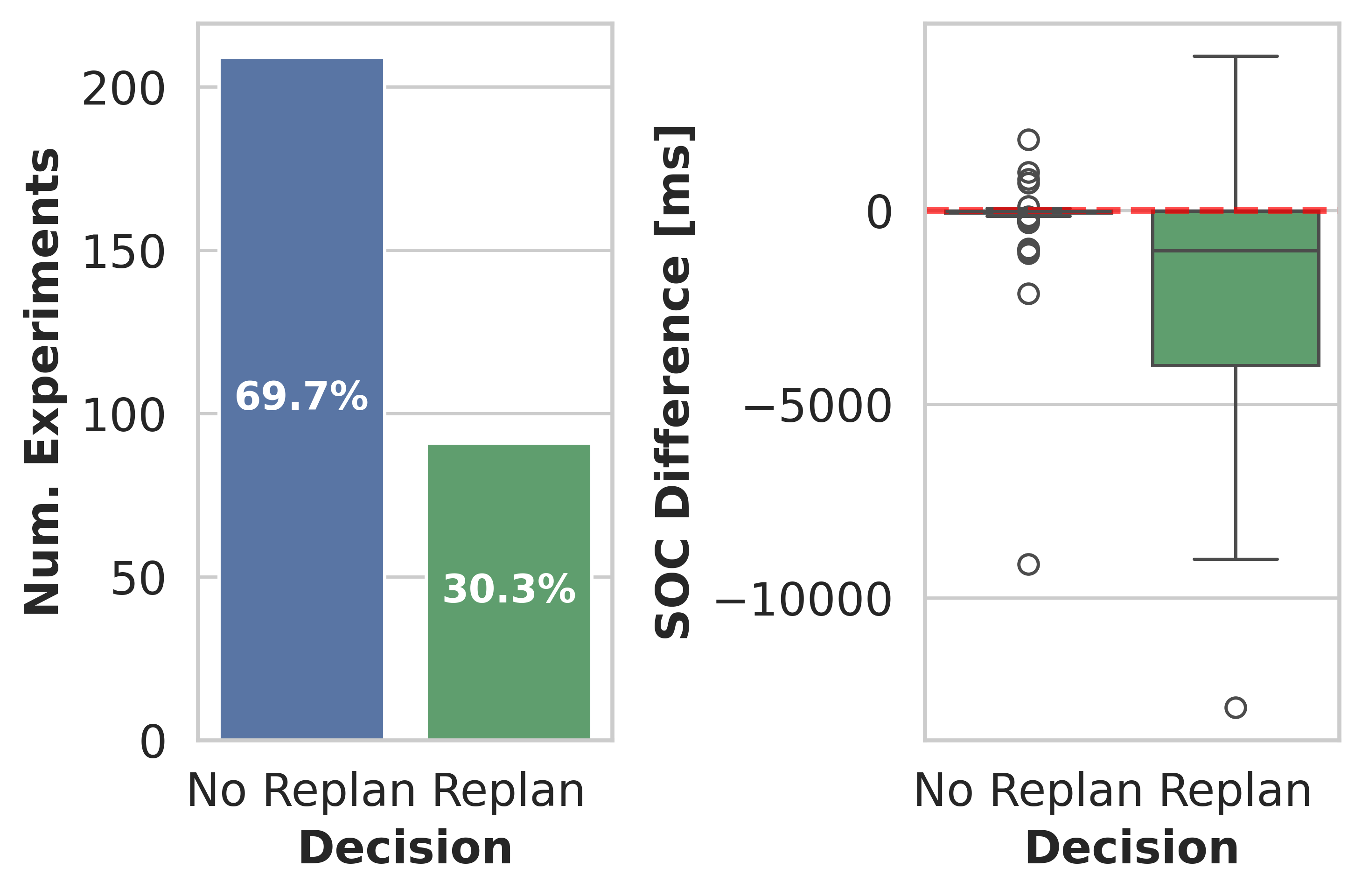}
        \caption{\texttt{random32}.}
        \label{subfig::impact_random32}
    \end{subfigure}
    \caption{Replanning frequency (left) and SOC Difference ($T_c^D = T_c^P - T_c^R$) (right) achieved with Slack-based replanning compared to SOC with Random replanning on different maps. Separated based on whether the Slack-based method decided to replan or not.}
    \label{fig:effect_of_replan_decision}
\end{figure}

Fig.~\ref{fig:effect_of_replan_decision} shows the behavior of predictive replanning and its effect on execution cost.
Because the original plans are optimal, replanning cannot improve the execution cost beyond minimizing the cost increase caused by the intruder.
Therefore, we show the difference $T_c^D = T_c^P - T_c^R$, where $T_c^P$ is the execution cost achieved with predictive replanning and $T_c^R$ is the execution cost after random replanning.
If $T_c^D = 0$, this means that both approaches resulted in the same execution cost.
If the predictive method did not replan, it is equivalent to a straightforward execution with the intruder's influence.

In approximately $\frac{1}{3}$ of the experiments, the predictive method decided to replan.
In these cases, the average SOC was lower than the SOC achieved while replanning randomly.
The exception is the \texttt{room32} map, where the average cost remained the same.
This shows that the predictive replanning method was capable of consistently triggering replanning at the right time.
In the other approximately $\frac{2}{3}$ of the experiments, the predictive method decided not to replan.
We can see that in such cases, the average execution cost was very close to the cost achieved by random replanning.
In other words, following both the original and the alternate plans resulted in very similar outcomes.
In combination, these results suggest that the slack-based predictive approach was able to determine whether replanning has the potential to improve execution cost or not.

There are two notable outliers on \texttt{arena} and \texttt{random32}, where following the original plan resulted in a noticably lower solution cost.
They can be seen in the No Replan column in Figs.~\ref{subfig::impact_arena} and~\ref{subfig::impact_random32}.
This is because the random method triggered replanning when the intruder was still present, but the new plan found by the replanning solver introduced a temporal dependency between the blocked agent and another agent.
This caused an increase in execution cost and performed worse than not replanning at all, even though the new plan was also optimal.

In some other cases, following the original plan could still lead to slightly better execution costs. 
However, this is a side effect of using real-time simulation and SOC as an objective.
When replanning is triggered, some agents may have to wait for another agent to finish its movement.
We consider this time spent waiting as part of SOC, and even a very short waiting time can accumulate to a noticeable increase in SOC if many agents are waiting at once.

\subsection{Mitigating the Intruder's Influence}
We computed the intruder's influence $I$ as the difference between the execution time without replanning $T_c^N$ and the lower bound of the execution time $T_c^{lb}$ (execution \emph{without} an intruder): $I = T_c^N - T_c^{lb}$.
Then, we computed the influence \emph{mitigated} by the random and predictive methods as $I^{R} = T_c^{R} - T_c^{lb}$ and $I^{P} = T_c^{P} - T_c^{lb}$, respectively.
Finally, we can express the \emph{mitigation} achieved by the methods as: $M^R = \frac{I^{R}}{I}$ and $M^P = \frac{I^{P}}{I}$.
The mitigation $M^P$ achieved by the predictive replanning on instances where it did replan was $25.27\%$ on \texttt{lab}, $27.55\%$ on \texttt{arena}, $27.70\%$ on \texttt{room32} and $28.95\%$ on \texttt{random32}. 
For random replanning, the mitigation $M^R$ achieved on the same instances was $3.45\%$ on \texttt{lab}, $7.07\%$ on \texttt{arena}, $10.92\%$ on \texttt{room32} and $9.81\%$ on \texttt{random32}.
The total average $\bar{M}^P=27.37\%$ and $\bar{M}^R=7.00\%$.
These values do not include outlier results which occurred due to the real-time nature of the simulation.
An example can be a situation where the predictive method did not replan, but achieved better execution cost; the two methods did replan at a similar time but achieved noticeably different results; or where a replanning method achieved better results than the lower bound of the execution.
Outliers were detected in $0.7\%$ of the\texttt{lab} experiments and in $4.3\%$ of the \texttt{arena} experiments.
On instances where the predictive method did not replan, the average mitigation was on average near zero for both predictive and random, suggesting that replanning had little impact.
These results show that the slack-based predictive approach was much more effective in reducing the impact of the intruder has on the execution cost than replanning randomly after the intruder appears.

\subsection{Influence of Fleet Size}

\begin{figure}[htb]
\centering
    \begin{subfigure}[c]{0.40\textwidth}
        \centering
        \hspace{-20pt}
        \includegraphics[width=1.0\textwidth]{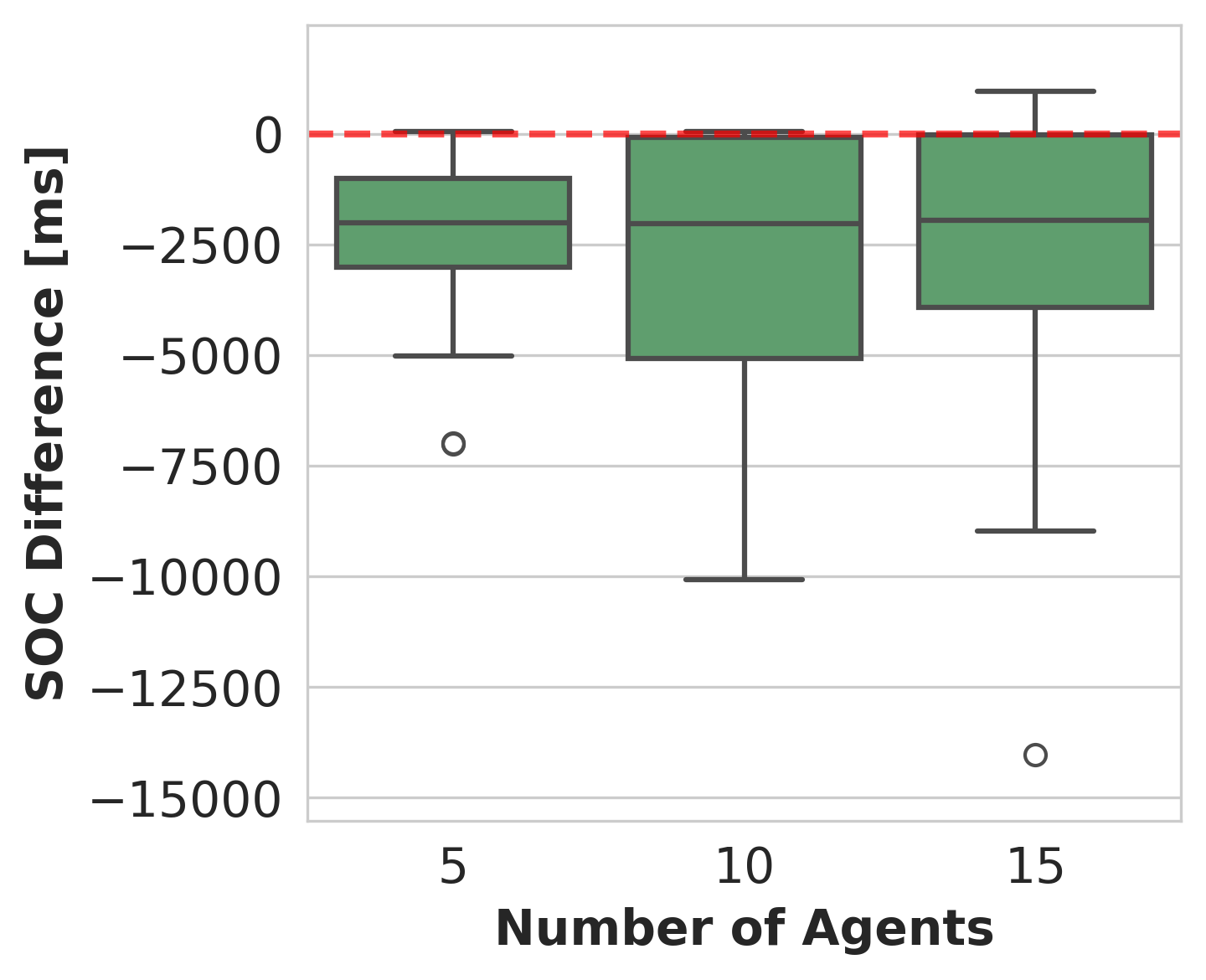}
        \caption{\texttt{lab}.}
    \end{subfigure}
    \begin{subfigure}[c]{0.40\textwidth}
        \centering
        \hspace{-20pt}
        \includegraphics[width=1.0\textwidth]{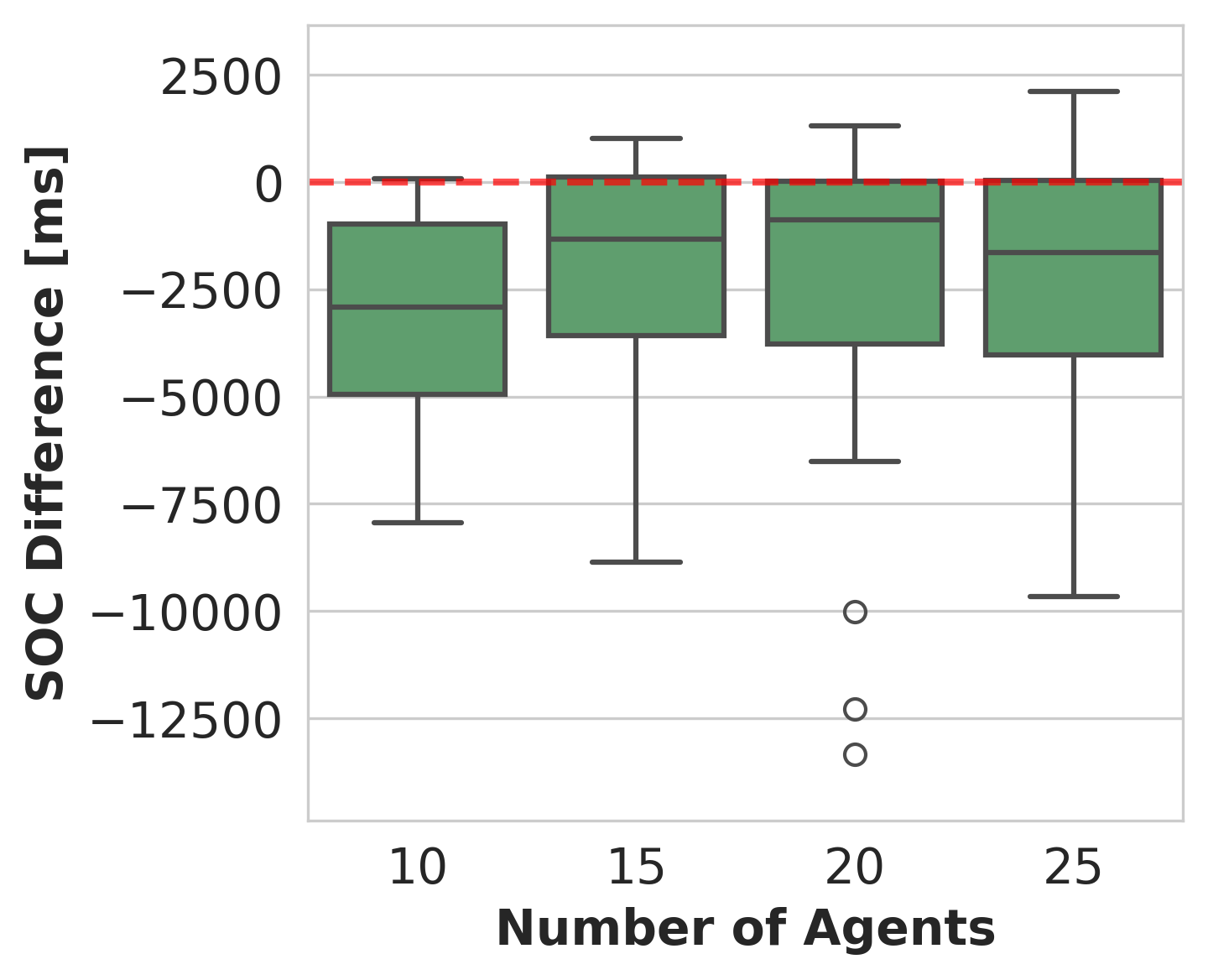}
        \caption{\texttt{arena}.}
    \end{subfigure}
    \begin{subfigure}[c]{0.40\textwidth}
        \centering
        \hspace{-20pt}
        \includegraphics[width=1.0\textwidth]{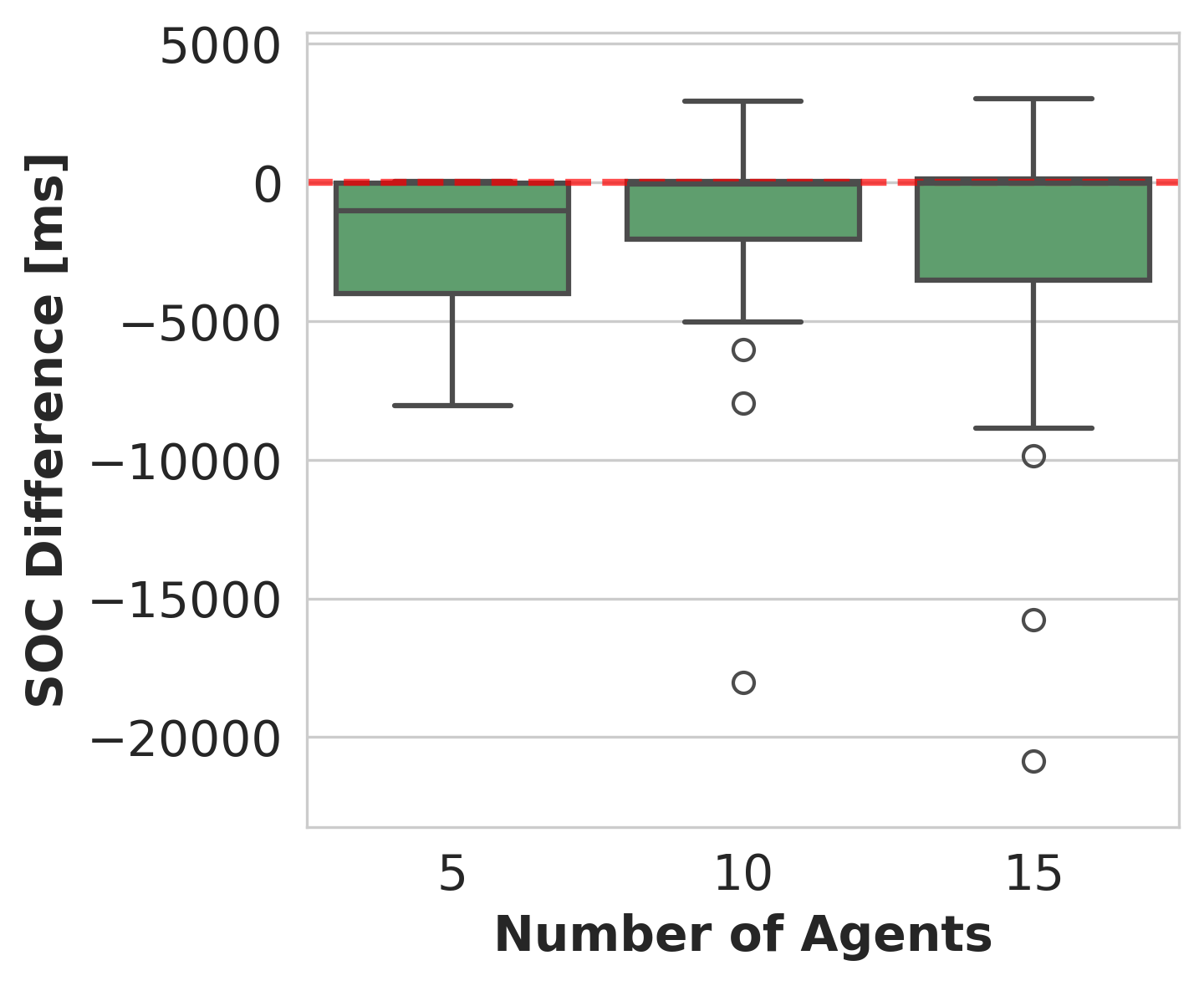}
        \caption{\texttt{room32}.}
    \end{subfigure}
    \begin{subfigure}[c]{0.40\textwidth}
        \centering
        \hspace{-20pt}
        \includegraphics[width=1.0\textwidth]{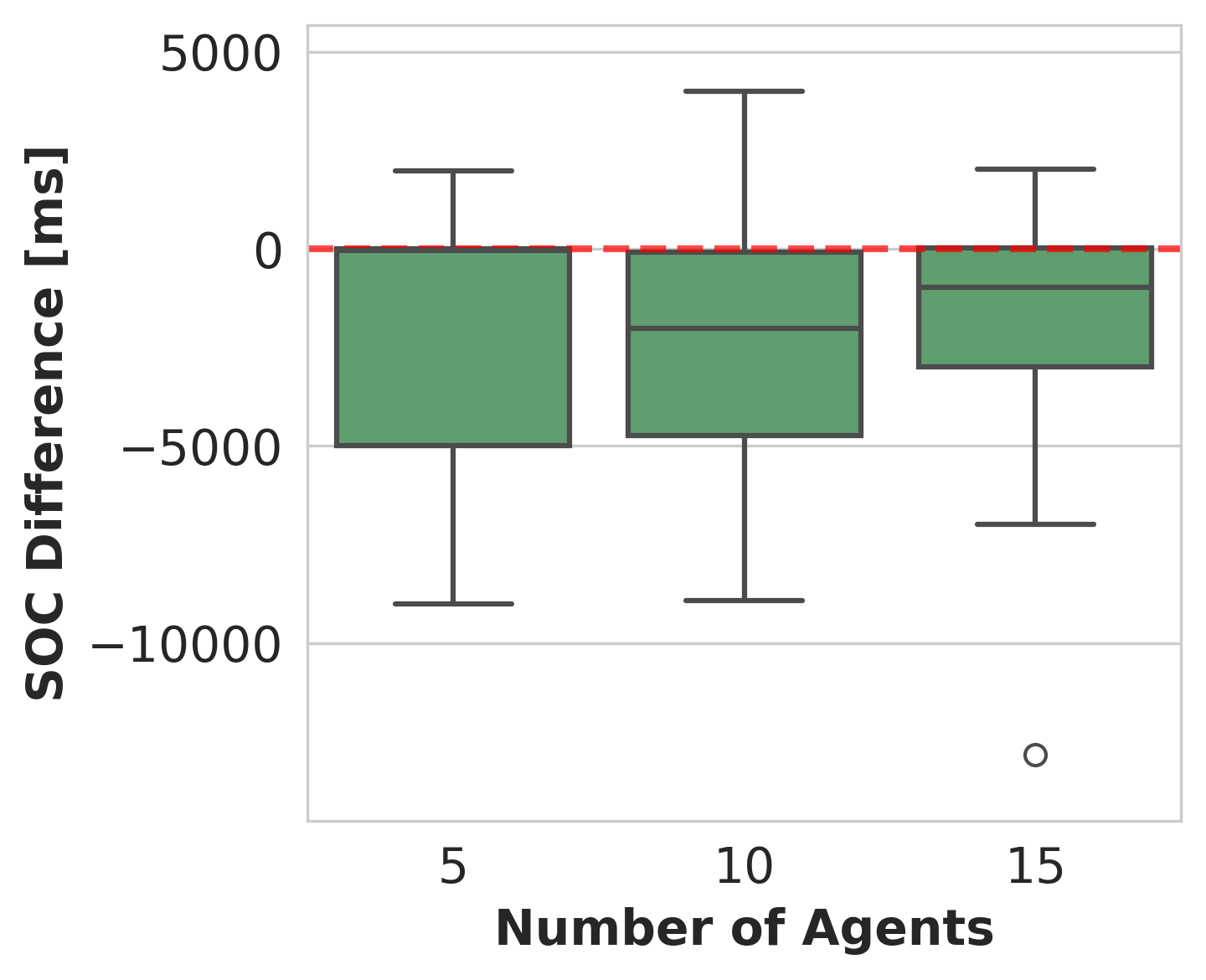}
        \caption{\texttt{random32}.}
    \end{subfigure}
    \caption{SOC Difference ($T_c^D = T_c^P - T_c^R$) achieved by predictive replanning on different maps for different numbers of agents. Only the instances where both methods replanned are shown. SOC is equivalent to milliseconds of execution.}
    \label{fig:effect_of_agents}
\end{figure}

Fig.~\ref{fig:effect_of_agents} shows how the number of agents influences the performance of the slack-based predictive approach compared to random replanning.
We can see that the number of agents did not have a significant impact on performance, except for instances with $10$ and $15$ agents on \texttt{room32} and with $5$ agents on \texttt{random32}, where the mean difference was close to zero.
However, even though the performance was similar, the predictive method still achieved significantly better results on some instances.

\subsection{Replanning Difficulty}
The average time to find an alternate solution by replanning was $\SI{8398}{\ms}$ for predictive replanning and $\SI{1774}{\ms}$ for random replanning.
Although the average is very high, the median values are much lower, with $\SI{80}{\ms}$ for predictive and $\SI{20}{\ms}$ for random.
This is because in very rare cases the fleet was at the time of replanning in a configuration that made it severely difficult to find a new plan.
In such outlier cases, the solver could take up to $\SI{90}{\s}$ to find a solution.

The discrepancy between the solution times for predictive and random replanning can be explained by the fact that random replanning times were normally distributed from the time the intruder appeared to near the end of the execution. 
Therefore, some agents may already have been in their goals, simplifying the task for the replanning solver.  
Slack-based replanning almost always occurred while most agents were still on their way to their goals.
With an increasing number of agents, the time required to find a new plan increases rapidly.

\section{Conclusion}
In this paper, we present an architecture that allows monitoring and optimizing the robust execution of MAPF plans.
We demonstrate the capability by implementing the architecture with a method that estimates the impact of delayed agents on the expected duration of the execution by computing the so-called \emph{slack}.  
To evaluate the method, we developed a simulator that closely mimics a real-life demonstrator of an autonomous robotic fleet for automated warehousing.
Delays were simulated using an unmodeled agent as an \emph{intruder}, which occupied a vertex in the shared operation space, blocking a single robot for a specified time.
We performed $1300$ experiments in total on four different maps with varying fleet sizes, in which we compare our developed method to a baseline method that replans at a random time after the intrusion.
The slack-based method was able to efficiently determine whether it is a good time to find an alternate plan by replanning.
It was able to reduce the impact of the intruder by $27.37\%$ on average compared to the average $7.00\%$ reduction achieved by the randomized replanning baseline.
Furthermore, it was able to distinguish between the cases where finding an alternate plan has the potential to reduce the execution cost and the cases where it would have little effect. 

%
%
%
\bibliographystyle{splncs04}
\bibliography{main}

\begin{thebibliography}{10}
\providecommand{\url}[1]{\texttt{#1}}
\providecommand{\urlprefix}{URL }
\providecommand{\doi}[1]{https://doi.org/#1}

\bibitem{Atzmon2020}
Atzmon, D., Stern, R., Felner, A., Wagner, G., Barták, R., Zhou, N.F.: Robust multi-agent path finding and executing. Journal of Artificial Intelligence Research  \textbf{67},  549--579 (3 2020). \doi{10.1613/jair.1.11734}, \url{https://jair.org/index.php/jair/article/view/11734}

\bibitem{barer2014suboptimal}
Barer, M., Sharon, G., Stern, R., Felner, A.: Suboptimal {{Variants}} of the {{Conflict-Based Search Algorithm}} for the {{Multi-Agent Pathfinding Problem}}. In: Seventh {{Annual Symposium}} on {{Combinatorial Search}} (Jul 2014)

\bibitem{Bartak}
Bart\'{a}k, R., Zhou, N.F., Stern, R., Boyarski, E., Surynek, P.: Modeling and {S}olving the {M}ulti-{A}gent {P}athfinding {P}roblem in {P}icat. In: IEEE 29th International Conference on Tools with Artificial Intelligence (ICTAI). pp. 959--966 (2017)

\bibitem{Berndt2020}
Berndt, A., Duijkeren, N.V., Palmieri, L., Keviczky, T.: A feedback scheme to reorder a multi-agent execution schedule by persistently optimizing a switchable action dependency graph. CoRR  \textbf{abs/2010.05254} (10 2020), \url{https://arxiv.org/abs/2010.05254v1}

\bibitem{Berndt2024}
Berndt, A., Duijkeren, N.V., Palmieri, L., Kleiner, A., Keviczky, T.: Receding horizon re-ordering of multi-agent execution schedules. IEEE Transactions on Robotics  \textbf{40},  1356--1372 (2024). \doi{10.1109/TRO.2023.3344051}

\bibitem{Chen2021}
Chen, Z., Harabor, D.D., Li, J., Stuckey, P.J.: Symmetry breaking for k-robust multi-agent path finding. Proceedings of the AAAI Conference on Artificial Intelligence  \textbf{35},  12267--12274 (5 2021). \doi{10.1609/aaai.v35i14.17456}, \url{https://ojs.aaai.org/index.php/AAAI/article/view/17456}

\bibitem{Dresner}
Dresner, K., Stone, P.: A {M}ultiagent {A}pproach to {A}utonomous {I}ntersection {M}anagement. Journal of Artificial Intelligence Research (JAIR)  \textbf{31},  591--656 (2008)

\bibitem{erdem2013general}
Erdem, E., Kisa, D.G., Oztok, U., Sch{\"u}ller, P.: A {{General Formal Framework}} for {{Pathfinding Problems}} with {{Multiple Agents}}. In: Twenty-{{Seventh AAAI Conference}} on {{Artificial Intelligence}} (Jun 2013)

\bibitem{Feng2024}
Feng, Y., Paul, A., Chen, Z., Li, J.: A real-time rescheduling algorithm for multi-robot plan execution. Proceedings of the International Conference on Automated Planning and Scheduling  \textbf{34},  201--209 (5 2024). \doi{10.1609/ICAPS.V34I1.31477}, \url{https://ojs.aaai.org/index.php/ICAPS/article/view/31477}

\bibitem{Gregoire2017}
Gregoire, J., Čáp, M., Frazzoli, E.: Locally-optimal multi-robot navigation under delaying disturbances using homotopy constraints. Autonomous Robots 2017 42:4  \textbf{42},  895--907 (12 2017). \doi{10.1007/S10514-017-9673-6}, \url{https://link.springer.com/article/10.1007/s10514-017-9673-6}

\bibitem{Honig2019}
H\"onig, W., Kiesel, S., Tinka, A., Durham, J.W., Ayanian, N.: Persistent and robust execution of mapf schedules in warehouses. IEEE Robotics and Automation Letters  \textbf{4},  1125--1131 (4 2019). \doi{10.1109/LRA.2019.2894217}

\bibitem{Hoenig2016}
H\"onig, W., Kumar, T.K., Cohen, L., Ma, H., Xu, H., Ayanian, N., Koenig, S.: Multi-agent path finding with kinematic constraints. Proceedings of the International Conference on Automated Planning and Scheduling  \textbf{26},  477--485 (3 2016). \doi{10.1609/ICAPS.V26I1.13796}, \url{https://ojs.aaai.org/index.php/ICAPS/article/view/13796}

\bibitem{Kottinger2024}
Kottinger, J., Geft, T., Almagor, S., Salzman, O., Lahijanian, M.: Introducing delays in multi-agent path finding. In: The International Symposium on Combinatorial Search (2024). \doi{10.1609/socs.v17i1.31540}

\bibitem{lehoux-lebacque2024multiagent}
Lehoux-Lebacque, V., Silander, T., Loiodice, C., Lee, S., Wang, A., Michel, S.: Multi-{Agent} {Path} {Finding} with {Real} {Robot} {Dynamics} and {Interdependent} {Tasks} for {Automated} {Warehouses}. In: {ECAI} 2024, pp. 4393--4401. IOS Press (2024). \doi{10.3233/FAIA241017}, \url{https://ebooks.iospress.nl/doi/10.3233/FAIA241017}

\bibitem{li2021anytime}
Li, J., Chen, Z., Harabor, D., Stuckey, P.J., Koenig, S.: Anytime {{Multi-Agent Path Finding}} via {{Large Neighborhood Search}}. In: Proceedings of the {{Thirtieth International Joint Conference}} on {{Artificial Intelligence}}. pp. 4127--4135. {International Joint Conferences on Artificial Intelligence Organization}, {Montreal, Canada} (Aug 2021). \doi{10.24963/ijcai.2021/568}

\bibitem{li2019symmetrybreaking}
Li, J., Harabor, D., Stuckey, P.J., Ma, H., Koenig, S.: Symmetry-{{Breaking Constraints}} for {{Grid-Based Multi-Agent Path Finding}}. Proceedings of the AAAI Conference on Artificial Intelligence  \textbf{33}(01),  6087--6095 (Jul 2019). \doi{10.1609/aaai.v33i01.33016087}

\bibitem{li2021eecbs}
Li, J., Ruml, W., Koenig, S.: {{EECBS}}: {{A Bounded-Suboptimal Search}} for {{Multi-Agent Path Finding}}. Proceedings of the AAAI Conference on Artificial Intelligence  \textbf{35}(14),  12353--12362 (May 2021)

\bibitem{Ma2017}
Ma, H., Kumar, T.K., Koenig, S.: Multi-agent path finding with delay probabilities. Proceedings of the AAAI Conference on Artificial Intelligence  \textbf{31},  3605--3612 (2 2017). \doi{10.1609/AAAI.V31I1.11035}, \url{https://ojs.aaai.org/index.php/AAAI/article/view/11035}

\bibitem{Morris}
Morris, R., Pasareanu, C., Luckow, K., Malik, W., Ma, H., Kumar, T., Koenig, S.: Planning, {S}cheduling and {M}onitoring for {A}irport {S}urface {O}perations. In: The Workshops of the Thirtieth AAAI Conference on Artificial Intelligence. pp. 608--614 (2016)

\bibitem{Okumura2023}
Okumura, K.: Improving lacam for scalable eventually optimal multi-agent pathfinding. In: Elkind, E. (ed.) Proceedings of the Thirty-Second International Joint Conference on Artificial Intelligence, {IJCAI-23}. pp. 243--251. International Joint Conferences on Artificial Intelligence Organization (8 2023). \doi{10.24963/ijcai.2023/28}, \url{https://doi.org/10.24963/ijcai.2023/28}, main Track

\bibitem{phillips2011sipp}
Phillips, M., Likhachev, M.: {{SIPP}}: {{Safe}} interval path planning for dynamic environments. In: 2011 {{IEEE International Conference}} on {{Robotics}} and {{Automation}}. pp. 5628--5635 (May 2011). \doi{10.1109/ICRA.2011.5980306}

\bibitem{sharon2015conflict}
Sharon, G., Stern, R., Felner, A., Sturtevant, N.R.: Conflict-based search for optimal multi-agent pathfinding. Artificial intelligence  \textbf{219},  40--66 (2015)

\bibitem{silver2005cooperative}
Silver, D.: Cooperative {{Pathfinding}}. Proceedings of the AAAI Conference on Artificial Intelligence and Interactive Digital Entertainment  \textbf{1}(1),  117--122 (2005). \doi{10.1609/aiide.v1i1.18726}

\bibitem{stern2019multi}
Stern, R., Sturtevant, N., Felner, A., Koenig, S., Ma, H., Walker, T., Li, J., Atzmon, D., Cohen, L., Kumar, T., et~al.: Multi-agent pathfinding: Definitions, variants, and benchmarks. In: Proceedings of the International Symposium on Combinatorial Search. vol.~10, pp. 151--158 (2019)

\bibitem{Su2024}
Su, Y., Veerapaneni, R., Li, J.: Bidirectional temporal plan graph: Enabling switchable passing orders for more efficient multi-agent path finding plan execution. Proceedings of the AAAI Conference on Artificial Intelligence  \textbf{38},  17559--17566 (3 2024). \doi{10.1609/AAAI.V38I16.29706}, \url{https://ojs.aaai.org/index.php/AAAI/article/view/29706}

\bibitem{surynek2010optimization}
Surynek, P.: An optimization variant of multi-robot path planning is intractable. In: Proceedings of the AAAI conference on artificial intelligence. vol.~24, pp. 1261--1263 (2010)

\bibitem{surynek2020bounded}
Surynek, P.: Bounded {{Sub-optimal Multi-Robot Path Planning Using Satisfiability Modulo Theory}} ({{SMT}}) {{Approach}}. In: 2020 {{IEEE}}/{{RSJ International Conference}} on {{Intelligent Robots}} and {{Systems}} ({{IROS}}). pp. 11631--11637. {IEEE}, {Las Vegas, NV, USA} (Oct 2020). \doi{10.1109/IROS45743.2020.9341047}

\bibitem{Wagner2022}
Wagner, A., Veerapaneni, R., Likhachev, M.: Minimizing coordination in multi-agent path finding with dynamic execution. Proceedings of the AAAI Conference on Artificial Intelligence and Interactive Digital Entertainment  \textbf{18},  61--69 (10 2022). \doi{10.1609/AIIDE.V18I1.21948}, \url{https://ojs.aaai.org/index.php/AIIDE/article/view/21948}

\bibitem{wang2016apriltaga}
Wang, J., Olson, E.: {AprilTag} 2: {Efficient} and robust fiducial detection. In: 2016 {IEEE}/{RSJ} {International} {Conference} on {Intelligent} {Robots} and {Systems} ({IROS}). pp. 4193--4198 (Oct 2016). \doi{10.1109/IROS.2016.7759617}, \url{https://ieeexplore.ieee.org/abstract/document/7759617}, iSSN: 2153-0866

\bibitem{Wu2024}
Wu, Y., Veerapaneni, R., Li, J., Likhachev, M.: From space-time to space-order: Directly planning a temporal planning graph by redefining cbs. arXiv  \textbf{2404.15137} (4 2024), \url{https://arxiv.org/abs/2404.15137v1}

\bibitem{Wurman}
Wurman, P.R., D'Andrea, R., Mountz, M., Mountz, M.: Coordinating {H}undreds of {C}ooperative, {A}utonomous {V}ehicles in {W}arehouses. AI Magazine  \textbf{29},  9--20 (03 2008)

\bibitem{JYU}
Yu, J., LaValle, S.M.: Optimal multirobot path planning on graphs: Complete algorithms and effective heuristics. IEEE Transactions on Robotics  \textbf{PP}(99),  1--15 (2016)

\bibitem{zahradka2023solving}
Zahr{\'a}dka, D., Kubi{\v{s}}ta, D., Kulich, M.: Solving robust execution of multi-agent pathfinding plans as a scheduling problem. Planning and Robotics, ICAPS'23 Workshop  (2023)

\bibitem{zahradka2025holistic}
Zahrádka, D., Mužíková, D., Kulich, M., Švancara, J., Barták, R.: Towards {Holistic} {Approach} to {Robust} {Execution} of {MAPF} {Plans}. In: Proceedings of the 17th {International} {Conference} on {Agents} and {Artificial} {Intelligence}. vol.~1, pp. 624--631. SCITEPRESS - Science and Technology Publications, Porto, Portugal (2025). \doi{10.5220/0013319700003890}, \url{https://www.scitepress.org/DigitalLibrary/Link.aspx?doi=10.5220/0013319700003890}

\end{thebibliography}
\end{document}